\documentclass[10pt]{article} 
\usepackage[left=0.75in, right=0.75in, bottom=1in]{geometry}

\usepackage{graphicx}  
\usepackage{amssymb}
\usepackage{multirow}
\usepackage{amsmath}
\usepackage{floatrow}
\usepackage[label font=bf,labelformat=simple]{subfig}
\usepackage{caption}
\usepackage{rotating}
\usepackage{array}
\usepackage{color} 
\usepackage{lineno}
\usepackage{comment}
\usepackage{natbib}
\usepackage{fancyhdr}
\usepackage{pbox}
\usepackage{tikz}
\newcommand{\boxalign}[2][0.97\textwidth]{
  \par\noindent\tikzstyle{mybox} = [draw=black,inner sep=6pt]
  \begin{center}\begin{tikzpicture}
   \node [mybox] (box){%
    \begin{minipage}{#1}{\vspace{-5mm}#2}\end{minipage}
   };
  \end{tikzpicture}\end{center}
}

   \usepackage{graphicx}
   \usepackage{url}
   \usepackage[colorlinks=false,urlcolor=blue]{hyperref}

   \usepackage{amsmath,amsthm,amsfonts,amssymb}
   \usepackage{multirow}
   \usepackage{floatflt}
   \usepackage{paralist}
   \usepackage[wide]{sidecap}
   \usepackage{enumitem}
   \usepackage{wrapfig}
   \usepackage[compact]{titlesec}
\DeclareMathOperator\erf{erf}

\setitemize[0]{leftmargin=*}

\def\text#1{{\rm #1}}

\def\vecx{\mathbf{x}}
\def\vecy{\mathbf{y}}

\def\<{\langle}
\def\>{\rangle}

\def\8{\infty}

\def\pip2{PIP$_2$}

\newcommand{\bt}[1]{ \begin{tabular} { #1 } }
\newcommand{\et} {\end{tabular}}
\newcommand{\ba}[1]{ \begin{array} { #1 } }
\newcommand{\ea} {\end{array}}
\newcommand{\beq}{ \begin{equation}}
\newcommand{\eeq}{\end{equation}}
\newcommand{\beqa}{\begin{eqnarray}}
\newcommand{\eeqa}{\end{eqnarray}}

\def\eps{\varepsilon}

\newcommand{\leavethisout} [1] {}

\newcommand{\bx}{{\mathbf x}}
\newcommand{\by}{{\mathbf y}}

\usepackage{authblk}

\title{Analytical steady-state solutions for pressure with a multiscale non-local model for two-fluid systems}
\author[1]{A. A. Howard}
\author[2]{Y. C. Zhou}
\author[1]{A. M. Tartakovsky}
\affil[1]{Pacific Northwest National Laboratory, Richland, WA}
\affil[2]{Department of Mathematics, Colorado
State University, Fort Collins, CO}
\date{}                     
\begin{document}
\maketitle


\begin{abstract}
We consider the nonlocal multiscale model for surface tension \citep{Tartakovsky2018} as an alternative to the (macroscale) Young-Laplace law. The nonlocal model is obtained in the form of an integral of a molecular-force-like function with support $\varepsilon$ added to the Navier-Stokes momentum conservation equation. Using this model, we calculate analytical forms for the steady-state equilibrium pressure gradient and pressure profile for circular and spherical bubbles and flat interfaces in two and three dimensions. 
According to the analytical solutions, the pressure changes continuously across the interface in a way that is quantitatively similar to what is observed in MD simulations. Furthermore, the pressure difference $P_{\varepsilon, in} -  P_{\varepsilon, out}$ satisfies the Young-Laplace law for the radius of curvature greater than $3\varepsilon$ and deviates from the Young-Laplace law otherwise (i.e., $P_{\varepsilon, in} -  P_{\varepsilon, out}$ goes to zero as the radius of the curvature goes to zero, where $P_{\varepsilon, out}$ is the pressure outside of the bubble at the distance greater than $3\varepsilon$ from the interface and  $P_{\varepsilon, in}$ is the pressure at the center of the bubble). The latter indicates that the surface tension in the proposed model decreases with the decreasing radius of curvature, which agrees with molecular dynamics simulations and laboratory experiments with nanobubbles. Therefore, our results demonstrate that the nonlocal model behaves microscopically at scales smaller than $\varepsilon$ and macroscopically, otherwise.  
\end{abstract}


\section{Non-local surface tension model}

We consider the case of two fluids, denoted $\alpha$ and $\beta$, in static equilibrium in a domain $\Omega = \Omega_\alpha \cup \Omega_\beta$. 
The fluid pressure $P_\alpha$ satisfies the static momentum conservation equation  \citep{Tartakovsky2018}:
\begin{equation}
\nabla P_\alpha = \mathbf{F} \quad\mathbf{x}\in\Omega_\alpha.
\label{Eq-Mom-Brackbill}
\end{equation}
where $\mathbf{F}$ is the nonlocal force due to surface tension  
\begin{equation}
\mathbf{F} = -\int_\Omega s(\mathbf{x}, \mathbf{y}) f_\varepsilon(|\mathbf{x}-\mathbf{y}|) \frac{\mathbf{x}-\mathbf{y}}{|\mathbf{x}-\mathbf{y}|} \; d\mathbf{y}, \; \; \mathbf{x} \in \Omega. 
\label{eq:Nonlocal_force}
\end{equation}
Here, $s(\mathbf{x}, \mathbf{y})$ is the force strength and $ f_\varepsilon(|\mathbf{x}-\mathbf{y}|)$ is the force shape function. 
The force strength is given by 
\begin{equation}
s(\mathbf{x}, \mathbf{y})= \left\{ \begin{array}{ll}
      s_{\alpha\alpha}, & \mathbf{x} \in \Omega_\alpha, \; \mathbf{y} \in \Omega_\alpha, \\
            s_{\alpha\beta}, & \mathbf{x} \in \Omega_\alpha, \; \mathbf{y} \in \Omega_\beta, \\
      s_{\beta\beta}, & \mathbf{x} \in \Omega_\beta, \; \mathbf{y} \in \Omega_\beta. \\
\end{array}
\right. 
\end{equation}
To ensure that $\sigma$ is positive, the coefficients must satisfy $  s_{\alpha\alpha} +   s_{\beta\beta} > 2  s_{\alpha\beta}$. Following \cite{Tartakovsky2016}, we take
$  s_{\alpha\alpha}  =   s_{\beta\beta}  =   10^k s_{\alpha\beta}$ with $k = 3$. Then, the coefficients are given by
\begin{equation}
 s_{\alpha\alpha}  =   s_{\beta\beta}  =   \frac{1}{2(1-10^{-k})}\frac{\sigma}{\lambda},
\end{equation}
where $\sigma$ is the macroscopic surface tension. The coefficient $\lambda$ depends on the shape of $f_\varepsilon$ as determined by \citet{Tartakovsky2016}: 
\begin{equation}
  \lambda =  \frac18 \pi  \int \limits_0^\infty z^4 
   f_{\varepsilon}(z)dz \label{eq:l2d}
\end{equation}
and
\begin{equation}
  \lambda = \frac13  \int \limits_0^\infty z^3   f_{\varepsilon}(z)dz \label{eq:l3d}
\end{equation}
in three and two spatial dimensions, respectively. 

The force shape function must be negative for small $|\mathbf{x}-\mathbf{y}|$ and positive for large $|\mathbf{x}-\mathbf{y}|$.
Several forms of $f_\varepsilon$ have been proposed in \citet{Tartakovsky2016}. In this paper we take
\begin{equation}
f_\varepsilon(|\mathbf{x}-\mathbf{y}|) = |\mathbf{x}-\mathbf{y}| \left[-A e^{-\frac{|\mathbf{x}-\mathbf{y}|^2}{2\varepsilon_0^2}}+ e^{-\frac{|\mathbf{x}-\mathbf{y}|^2}{2\varepsilon^2}}\right]. \label{eq:forceshape}
\end{equation}
For $f_\varepsilon$ given by eq. \ref{eq:forceshape}, eq. \ref{eq:l2d} gives $\lambda = \frac{1}{2} \sqrt{2\pi}\left(\varepsilon^5-A\varepsilon_0^5\right)$ in two dimensions and eq. \ref{eq:l3d} gives $\lambda = \pi\left(\varepsilon^6-A\varepsilon_0^6\right)$ in three dimensions. For the figures in this paper we use $A =\left( \frac{\varepsilon}{\varepsilon_0}\right)^3$ and $\varepsilon_0 = 0.5\varepsilon$, although the analytical results are independent of the choice of $A$ and $\varepsilon_0$. In numerical simulations the parameter $\varepsilon$ is picked to be on the order of the grid spacing.

We note that the Young-Laplace (YL) law
\begin{equation}\label{Young-Laplace}
P_\alpha-P_\beta = \kappa \sigma 
\end{equation}
results in eq. \ref{Eq-Mom-Brackbill} with 
 \begin{equation}
\mathbf{F} = \sigma \kappa \nabla \phi,
\label{eq:YLlaw}
\end{equation}
where $\kappa$ is the interface curvature and $\phi$ is the color function 
\begin{equation}
\phi(\mathbf{x}) = \left\{ \begin{array}{ll}
      0, & \mathbf{x} \in \Omega_\alpha, \\
       1, & \mathbf{x} \in \Omega_\beta.
\end{array}
\right. 
\label{equ:color_func}
\end{equation}
 
\section{Two-dimensional bubble}

Consider a circular bubble with radius $a$ centered at $(0, 0)$. In steady state, we have: 
\begin{equation}
\nabla P(\bx) = -\int s(\bx, \by) f_\varepsilon(\bx - \by)\frac{\bx-\by}{|\bx-\by|} \; d\by = 
- \int s(\bx, \by) \left[e^{\frac{-|\bx-\by|^2}{2\varepsilon^2}}-Ae^{\frac{-|\bx-\by|^2}{2\varepsilon_0^2}}\right](\bx-\by) \; d\by \label{eq:2dequation}
\end{equation}

For convenience, we work in polar coordinates. Let $\by = (r \cos\theta, r\sin\theta)$ and $\bx = (r_i \cos\theta_i, r_i\sin\theta_i)$. Then, $\bx-\by = (r_i \cos\theta_i-r\cos\theta, r_i \sin\theta_i-r\sin\theta)^T$ and 
$|\bx-\by|^2 = r_i^2 + r^2 -2rr_i(\cos\theta_i\cos\theta+\sin\theta_i\sin\theta) =  r_i^2 + r^2 -2rr_i(\cos(\theta-\theta_i))$. We get: 

\begin{align}
\frac{\partial P}{\partial r_i} &=- \int_0^\infty \int_0^{2\pi} s(\bx, \by) \left[e^{\frac{-(r^2 + r_i^2 -2rr_i\cos(\theta-\theta_i))}{2\varepsilon^2}}-Ae^{\frac{-(r^2 + r_i^2 -2rr_i\cos(\theta-\theta_i))}{2\varepsilon_0^2}}\right]  \left[ (r_i\cos\theta_i -r\cos\theta)\cos\theta_i \right.\nonumber  \\
&\; \; \; +\left. (r_i\sin\theta_i -r\sin\theta)\sin\theta_i\right]
r  \; d\theta dr   \\
&\; \nonumber \\
\frac{1}{r_i}\frac{\partial P}{\partial \theta_i} &= -\int_0^\infty \int_0^{2\pi} s(\bx, \by) \left[e^{\frac{-(r^2 + r_i^2 -2rr_i\cos(\theta-\theta_i))}{2\varepsilon^2}}-Ae^{\frac{-(r^2 + r_i^2 -2rr_i\cos(\theta-\theta_i))}{2\varepsilon_0^2}}\right]  \left[ -(r_i\cos\theta_i -r\cos\theta)\sin\theta_i \right. \nonumber\\
&\; \; \; +\left. (r_i\sin\theta_i -r\sin\theta)\cos\theta_i\right]
r  \; d\theta dr 
\end{align}

\noindent First consider $\frac{\partial P}{\partial \theta}:$ 
\begin{align}
\frac{1}{r_i}\frac{\partial P}{\partial \theta_i} &= -\int_0^\infty \int_0^{2\pi} s(\bx, \by) \left[e^{\frac{-(r^2 + r_i^2 -2rr_i\cos(\theta-\theta_i))}{2\varepsilon^2}}-Ae^{\frac{-(r^2 + r_i^2 -2rr_i\cos(\theta-\theta_i))}{2\varepsilon_0^2}}\right]  \left[ -(r_i\cos\theta_i -r\cos\theta)\sin\theta_i \right. \nonumber\\
&\; \; \; +\left. (r_i\sin\theta_i -r\sin\theta)\cos\theta_i\right]
r  \; d\theta dr   \nonumber \\
&= \int_0^\infty \int_0^{2\pi} s(\bx, \by) \left[e^{\frac{-(r^2 + r_i^2 -2rr_i\cos(\theta-\theta_i))}{2\varepsilon^2}}-Ae^{\frac{-(r^2 + r_i^2 -2rr_i\cos(\theta-\theta_i))}{2\varepsilon_0^2}}\right] 
r^2 \sin(\theta-\theta_i)  \; d\theta dr   \nonumber\\
 & =  0.
\end{align}

\noindent Now we turn to $\frac{\partial P}{\partial r_i} $: 
\begin{align}
\frac{\partial P}{\partial r_i} &=- \int_0^\infty \int_0^{2\pi} s(\bx, \by) \left[e^{\frac{-(r^2 + r_i^2 -2rr_i\cos(\theta-\theta_i))}{2\varepsilon^2}}-Ae^{\frac{-(r^2 + r_i^2 -2rr_i\cos(\theta-\theta_i))}{2\varepsilon_0^2}}\right]  \left[ (r_i\cos\theta_i -r\cos\theta)\cos\theta_i \right.\nonumber  \\
&\; \; \; +\left. (r_i\sin\theta_i -r\sin\theta)\sin\theta_i\right]
r  \; d\theta dr   \nonumber \\
& = - \int_0^\infty \int_0^{2\pi} s(\bx, \by) \left[e^{\frac{-(r^2 + r_i^2 -2rr_i\cos(\theta-\theta_i))}{2\varepsilon^2}}-Ae^{\frac{-(r^2 + r_i^2 -2rr_i\cos(\theta-\theta_i))}{2\varepsilon_0^2}}\right] \left[r_i - r\cos(\theta-\theta_i) \right]
r  \; d\theta dr  
\end{align}
We will consider the integral for $\kappa$ where $\kappa = 2\varepsilon^2$ or $\kappa = 2\varepsilon_0^2$ on an interval $[b, c]$ where $ s(\bx, \by)$ is constant (i.e., $[b, c] \in [0,a)$ or $[b, c] \in (a, \infty)$.)
\begin{align}
& \int_b^c \int_0^{2\pi} s(\bx, \by) e^{\frac{-(r^2 + r_i^2 -2rr_i\cos(\theta-\theta_i))}{2\varepsilon^2}} (r_i -r\cos(\theta-\theta_i))
r  \; d\theta dr  =   2\pi s  \int_b^c   e^{\frac{-(r^2 + r_i^2 )}{\kappa}} \left[  rr_i  I_0\left(\frac{2rr_i}{\kappa}\right)-r^2  I_1\left(\frac{2rr_i}{\kappa}\right) \right]dr  \label{eq:one}
\end{align}
Now consider the first term in eq. \ref{eq:one} and substitute $w  = \frac{2rr_i}{\kappa}$:
\begin{equation}
2\pi s e^{-\frac{r_i^2 }{\kappa}} \int_b^c   e^{-\frac{r^2 }{\kappa}} rr_i  I_0\left(\frac{2rr_i}{\kappa}\right)\; dr = 
 \frac{2\pi s \kappa^2 }{4 r_i} e^{-\frac{r_i^2 }{\kappa}} \int_{2br_i/\kappa}^{2cr_i/\kappa}   e^{-\frac{w^2 }{\tilde{\kappa}}} w  I_0\left(w \right)\; dw
\end{equation}
where $\tilde{\kappa} = (2r_i)^2/\kappa.$ We integrate by parts with $u =  e^{-\frac{w^2 }{\tilde{\kappa}}}$ and $dv = w  I_0\left(w \right)\; dw$. 
This gives
\begin{align}
 \frac{2\pi s \kappa^2 }{4 r_i} e^{-\frac{r_i^2 }{\kappa}} \int_{2br_i/\kappa}^{2cr_i/\kappa}   e^{-\frac{w^2 }{\tilde{\kappa}}} w  I_0\left(w \right)\; dw  
 &=  \frac{2\pi s \kappa^2 }{4 r_i} e^{-\frac{r_i^2 }{\kappa}} \left\{   \left[ e^{-\frac{w^2 }{\tilde{\kappa}}} w  I_1\left(w \right)  \right]_{2br_i/\kappa}^{2cr_i/\kappa}
  + \frac{2}{\tilde{\kappa}} \int_{2br_i/\kappa}^{2cr_i/\kappa}  w^2 e^{-\frac{w^2 }{\tilde{\kappa}}}  I_1\left(w \right)\; dw
  \right\} \nonumber \\
 &= 
2\pi s  e^{-\frac{r_i^2 }{\kappa}} \left\{   \left[   \frac{ \kappa}{2 }  e^{-\frac{r^2 }{\kappa}}  r I_1\left( \frac{2rr_i}{\kappa} \right)  \right]_b^c
 + \int_b^c r^2 e^{-\frac{r^2 }{\kappa}}  I_1\left( \frac{2rr_i}{\kappa} \right)\; dr \label{eq:two}
  \right\}
\end{align}

Therefore, substituting eq. \ref{eq:two} into eq. \ref{eq:one} we have: 
\begin{align}
 & 2\pi s  \int_b^c   e^{\frac{-(r^2 + r_i^2 )}{\kappa}} \left[  rr_i  I_0\left(\frac{2rr_i}{\kappa}\right)-r^2  I_1\left(\frac{2rr_i}{\kappa}\right) \right]dr  \nonumber \\
& =   2\pi s  e^{-\frac{r_i^2 }{\kappa}} \left\{   \left[   \frac{ \kappa}{2 }  e^{-\frac{r^2 }{\kappa}}  r I_1\left( \frac{2rr_i}{\kappa} \right)  \right]_b^c
 + \int_b^c r^2 e^{-\frac{r^2 }{\kappa}}  I_1\left( \frac{2rr_i}{\kappa} \right)\; dr
-   \int_b^c   e^{-\frac{r^2  }{\kappa}} r^2  I_1\left(\frac{2rr_i}{\kappa}\right)dr    \right\} \nonumber \\
& =   2\pi s  e^{-\frac{r_i^2 }{\kappa}}  \left[   \frac{ \kappa}{2 }  e^{-\frac{r^2 }{\kappa}}  r I_1\left( \frac{2rr_i}{\kappa} \right)  \right]_b^c  
\end{align}

If $r_i \leq a$, $ s(r, r_i) = s_{\alpha \alpha}$ if $r \leq a$ and $ s(r, r_i) = s_{ab}$ if $r > a$. Therefore, 
\begin{align}
& 2\pi   \int_0^\infty s(r, r_i)   e^{\frac{-(r^2 + r_i^2 )}{\kappa}} \left[  rr_i  I_0\left(\frac{2rr_i}{\kappa}\right)-r^2  I_1\left(\frac{2rr_i}{\kappa}\right) \right]dr  \nonumber \\
&= 2\pi   \int_0^a s_{\alpha \alpha}   e^{\frac{-(r^2 + r_i^2 )}{\kappa}} \left[  rr_i  I_0\left(\frac{2rr_i}{\kappa}\right)-r^2  I_1\left(\frac{2rr_i}{\kappa}\right) \right]dr + 2\pi   \int_a^\infty s_{ab}   e^{\frac{-(r^2 + r_i^2 )}{\kappa}} \left[  rr_i  I_0\left(\frac{2rr_i}{\kappa}\right)-r^2  I_1\left(\frac{2rr_i}{\kappa}\right) \right]dr \nonumber  \\
&= 2\pi s_{\alpha \alpha}  e^{-\frac{r_i^2 }{\kappa}}  \left[   \frac{ \kappa}{2 }  e^{-\frac{r^2 }{\kappa}}  r I_1\left( \frac{2rr_i}{\kappa} \right)  \right]_0^a   + 2\pi s_{ab}  e^{-\frac{r_i^2 }{\kappa}}  \left[   \frac{ \kappa}{2 }  e^{-\frac{r^2 }{\kappa}}  r I_1\left( \frac{2rr_i}{\kappa} \right)  \right]_a^\infty\nonumber \\
&= 2\pi s_{\alpha \alpha}  e^{-\frac{r_i^2 }{\kappa}}  \left[   \frac{ \kappa}{2 }  e^{-\frac{a^2 }{\kappa}}  a I_1\left( \frac{2ar_i}{\kappa} \right)  \right]   - 2\pi s_{ab}  e^{-\frac{r_i^2 }{\kappa}}  \left[   \frac{ \kappa}{2 }  e^{-\frac{a^2 }{\kappa}}  a I_1\left( \frac{2ar_i}{\kappa} \right)  \right] \nonumber \\
&= 
a \kappa \pi( s_{\alpha \alpha} -s_{\alpha \beta}) e^{-\frac{r_i^2+a^2 }{\kappa}}    I_1\left( \frac{2ar_i}{\kappa} \right) 
\end{align}
A similar expression will hold if $r_i > a$, with $s_{\alpha \alpha}$ and $s_{\alpha \beta}$ switched. 

Now, 
\begin{align}
\frac{\partial P}{\partial r_i} &= -\int_0^\infty \int_0^{2\pi} s(\bx, \by) \left[e^{\frac{-(r^2 + r_i^2 -2rr_i\cos\theta)}{2\varepsilon^2}}-Ae^{\frac{-(r^2 + r_i^2 -2rr_i\cos\theta)}{2\varepsilon_0^2}}\right] (r_i -r\cos\theta)
r  \; d\theta dr  \nonumber \\
& = 
-2a\varepsilon^2 \pi( s_{\alpha \alpha} -s_{\alpha \beta}) e^{-\frac{r_i^2+a^2 }{2\varepsilon^2}}    I_1\left( \frac{ar_i}{\varepsilon^2} \right)-2Aa\varepsilon_0^2 \pi( s_{\alpha \alpha} -s_{\alpha \beta}) e^{-\frac{r_i^2+a^2 }{2\varepsilon_0^2}}    I_1\left( \frac{ar_i}{\varepsilon_0^2} \right) \nonumber \\
& = 
-2a\pi( s_{\alpha \alpha} -s_{\alpha \beta}) \left[\varepsilon^2  e^{-\frac{r_i^2+a^2 }{2\varepsilon^2}}    I_1\left( \frac{ar_i}{\varepsilon^2} \right)-A\varepsilon_0^2 e^{-\frac{r_i^2+a^2 }{2\varepsilon_0^2}}    I_1\left( \frac{ar_i}{\varepsilon_0^2} \right)\right]
\end{align}

\noindent To summarize, 
\begin{align}
\frac{\partial P}{\partial r_i} &= -2a\pi( s_{\alpha \alpha} -s_{\alpha \beta}) \left[\varepsilon^2  e^{-\frac{r_i^2+a^2 }{2\varepsilon^2}}    I_1\left( \frac{ar_i}{\varepsilon^2} \right)-A\varepsilon_0^2 e^{-\frac{r_i^2+a^2 }{2\varepsilon_0^2}}    I_1\left( \frac{ar_i}{\varepsilon_0^2} \right)\right] \\
\frac{\partial P}{\partial \theta_i} &= 0
\end{align}

Switching back to Cartesian coordinates, 
\begin{align}
\frac{\partial P}{\partial x} &= -2a\pi( s_{\alpha \alpha} -s_{\alpha \beta}) \left[\varepsilon^2  e^{-\frac{|\bx|^2+a^2 }{2\varepsilon^2}}    I_1\left( \frac{a|\bx|}{\varepsilon^2} \right)-A\varepsilon_0^2 e^{-\frac{|\bx|^2+a^2 }{2\varepsilon_0^2}}    I_1\left( \frac{a|\bx|}{\varepsilon_0^2} \right)\right] \frac{x}{|\bx|}\label{eq:dpdxhalf} \\
\frac{\partial P}{\partial y} &= -2a\pi( s_{\alpha \alpha} -s_{\alpha \beta}) \left[\varepsilon^2  e^{-\frac{|\bx|^2+a^2 }{2\varepsilon^2}}    I_1\left( \frac{a|\bx|}{\varepsilon^2} \right)-A\varepsilon_0^2 e^{-\frac{|\bx|^2+a^2 }{2\varepsilon_0^2}}    I_1\left( \frac{a|\bx|}{\varepsilon_0^2} \right)\right]\frac{y}{|\bx|} 
\label{eq:dpdyhalf}
\end{align}

Equations \ref{eq:dpdxhalf} and \ref{eq:dpdyhalf}  hold if $r = |\bx| < a$. If $r = |\bx| > a$ they must be multiplied by -1. Introduce the Heaviside step function $H$, given by 
\[  H[x] = \begin{cases} 
      0 & x\leq 0 \\
      1 & x > 0.
   \end{cases}
\]
Then, 
\boxalign{\begin{align}
\frac{\partial P}{\partial x} &= -(1-2H[|\bx| -a])2a\pi( s_{\alpha \alpha} -s_{\alpha \beta}) \left[\varepsilon^2  e^{-\frac{|\bx|^2+a^2 }{2\varepsilon^2}}    I_1\left( \frac{a|\bx|}{\varepsilon^2} \right)-A\varepsilon_0^2 e^{-\frac{|\bx|^2+a^2 }{2\varepsilon_0^2}}    I_1\left( \frac{a|\bx|}{\varepsilon_0^2} \right)\right] \frac{x}{|\bx|}\label{eq:dpdx} \\
\frac{\partial P}{\partial y} &= -(1-2H[|\bx|-a] )2a\pi( s_{\alpha \alpha} -s_{\alpha \beta}) \left[\varepsilon^2  e^{-\frac{|\bx|^2+a^2 }{2\varepsilon^2}}    I_1\left( \frac{a|\bx|}{\varepsilon^2} \right)-A\varepsilon_0^2 e^{-\frac{|\bx|^2+a^2 }{2\varepsilon_0^2}}    I_1\left( \frac{a|\bx|}{\varepsilon_0^2} \right)\right]\frac{y}{|\bx|}\label{eq:dpdy}
\end{align}}

To find the pressure we need to integrate eq. \ref{eq:dpdx}: 
\begin{equation}
P(x) - P(\infty) = -\int_\infty^x  (1-2H[|\bx_i|-a ])2a\pi( s_{\alpha \alpha} -s_{\alpha \beta}) \left[\varepsilon^2  e^{-\frac{|\bx_i|^2+a^2 }{2\varepsilon^2}}    I_1\left( \frac{a|\bx_i|}{\varepsilon^2} \right)-A\varepsilon_0^2 e^{-\frac{|\bx_i|^2+a^2 }{2\varepsilon_0^2}}    I_1\left( \frac{a|\bx_i|}{\varepsilon_0^2} \right)\right] \frac{x_i}{|\bx_i|}\; dx_i \label{eq:three}
\end{equation}

Note that the modified Bessel function of the first kind can be represented as an infinite sum of polynomials, with $u = \sqrt{x_i^2+y_i^2}$. 
\begin{align}
&\int_x^\infty \left(1-2H[\sqrt{x_i^2+y_i^2}-a ]\right) e^{-\frac{x_i^2+y_i^2}{2\varepsilon^2}}    I_1\left( \frac{a\sqrt{x_i^2+y_i^2}}{\varepsilon^2} \right)\frac{x_i}{\sqrt{x_i^2+y_i^2}}  \; dx_i \nonumber \\
& =
\int_{\sqrt{x^2+y_i^2}}^\infty \left(1-2H[u-a ]\right) e^{-\frac{u^2}{2\varepsilon^2}}    \sum_{l = 0}^\infty \frac{1}{l!(l+1)!} \left( \frac{au}{\varepsilon^2} \right)^{2l+1} \; du \nonumber \\
  & =  \sum_{l = 0}^\infty \frac{a}{l!(l+1)!} \left( \frac{a^2}{2\varepsilon^2} \right)^{l}  \left[-\frac{1}{2}\Gamma\left(l+1,  \frac{u^2}{2\varepsilon^2}\right)   + H[u-a ]\left(  \Gamma\left(l+1,  \frac{a^2}{2\varepsilon^2}\right)  -\Gamma\left(l+1,  \frac{u^2}{2\varepsilon^2}\right)         \right) \right]_{\sqrt{x^2+y_i^2}}^\infty \nonumber \\
    & =  \sum_{l = 0}^\infty \frac{a}{l!(l+1)!} \left( \frac{a^2}{2\varepsilon^2} \right)^{l}  \left[-\frac{1}{2}\Gamma\left(l+1,  \frac{x_i^2+y_i^2}{2\varepsilon^2}\right)   + H\left[\sqrt{x_i^2+y_i^2}-a \right]\left(  \Gamma\left(l+1,  \frac{a^2}{2\varepsilon^2}\right)  -\Gamma\left(l+1,  \frac{x_i^2+y_i^2}{2\varepsilon^2}\right)         \right) \right]_{x}^\infty \nonumber \\
  &=   \sum_{l = 0}^\infty \frac{a}{l!(l+1)!} \left( \frac{a^2}{2\varepsilon^2} \right)^{l}   \Gamma\left(l+1,  \frac{a^2}{2\varepsilon^2}\right)\nonumber \\
  & \quad -\sum_{l = 0}^\infty \frac{a}{l!(l+1)!} \left( \frac{a^2}{2\varepsilon^2} \right)^{l}  \left[\frac{1}{2}\Gamma\left(l+1,  \frac{x^2+y_i^2}{2\varepsilon^2}\right)   + H\left[\sqrt{x^2+y_i^2}-a \right]\left(  \Gamma\left(l+1,  \frac{a^2}{2\varepsilon^2}\right)  -\Gamma\left(l+1,  \frac{x^2+y_i^2}{2\varepsilon^2}\right)         \right) \right]  \label{eq:five}
\end{align}
\noindent where we used the fact that $\Gamma(k, x)$ is the upper incomplete gamma function, which has the properties $\Gamma(k, 0) = \Gamma(k) = (k-1)!$ and $\lim_{x \rightarrow \infty} \Gamma(k, x) = 0$. Also, $H[0-a] = 0$ and $\lim_{x \rightarrow \infty} H[x-a] = 1$. Substituting eq. \ref{eq:five} in to eq. \ref{eq:three} gives the pressure at the center of the bubble. We can use a similar process to find the pressure at any point $r = \sqrt{x^2+y^2}$: 
\begin{equation}
\boxed{P(r)-P(\infty) 
= 4 \pi( s_{\alpha \alpha} -s_{\alpha \beta}) \left(\varepsilon^4 G(r, \varepsilon) - A\varepsilon_0^4 G(r, \varepsilon_0)\right)} \label{eq:2dpressure}
\end{equation}
where 
\begin{equation}
G(r, \varepsilon)  = \begin{cases} e^{-\frac{a^2}{2\varepsilon^2}} \sum_{l = 0}^\infty \frac{1}{l!(l+1)!} \left( \frac{a^2}{2\varepsilon^2} \right)^{l+1}  \left[\Gamma\left(l+1,  \frac{a^2}{2\varepsilon^2}\right) -\frac{1}{2}\Gamma\left(l+1,  \frac{r^2}{2\varepsilon^2}\right) \right] & r < a \\
 e^{-\frac{a^2}{2\varepsilon^2}} \sum_{l = 0}^\infty \frac{1}{l!(l+1)!} \left( \frac{a^2}{2\varepsilon^2} \right)^{l+1}  \left[\frac{1}{2}\Gamma\left(l+1,  \frac{r^2}{2\varepsilon^2}\right) \right] & r \geq a \end{cases}
\end{equation}
It is easy to check that this is continuous at $r = a$. 

Note that $e^{-\frac{a^2}{2\varepsilon^2}} \sum_{l = 0}^\infty \frac{1}{l!(l+1)!} \left( \frac{a^2}{2\varepsilon^2} \right)^{l+1}  \Gamma\left(l+1,  \frac{r^2}{2\varepsilon^2}\right)$  can be represented as: 
\begin{equation}
e^{-\frac{a^2}{2\varepsilon^2}} \sum_{l = 0}^\infty \frac{1}{l!(l+1)!} \left( \frac{a^2}{2\varepsilon^2} \right)^{l+1} \Gamma\left(l+1,  \frac{r^2}{2\varepsilon^2}\right) = e^{-\frac{a^2+r^2}{2\varepsilon^2}} \left[\Phi_3\left(1, 1, \frac{a^2}{2\varepsilon^2}, \frac{a^2r^2}{4\varepsilon^4}\right) - I_0\left( \frac{ar}{\varepsilon^2}\right)  \right]
\end{equation}
where $\Phi_3$ denotes the Humbert series and $I_0$ is a modified Bessel function of the first kind. Therefore, 
\begin{equation}
G(r, \varepsilon)  = \begin{cases} 
e^{-\frac{a^2}{\varepsilon^2}} \left[\Phi_3\left(1, 1, \frac{a^2}{2\varepsilon^2}, \frac{a^4}{4\varepsilon^4}\right) - I_0\left( \frac{a^2}{\varepsilon^2}\right)  \right]
 - \frac{1}{2}e^{-\frac{a^2+r^2}{2\varepsilon^2}} \left[\Phi_3\left(1, 1, \frac{a^2}{2\varepsilon^2}, \frac{a^2r^2}{4\varepsilon^4}\right) - I_0\left( \frac{ar}{\varepsilon^2}\right)  \right] & r < a \\
 \frac{1}{2}e^{-\frac{a^2+r^2}{2\varepsilon^2}} \left[\Phi_3\left(1, 1, \frac{a^2}{2\varepsilon^2}, \frac{a^2r^2}{4\varepsilon^4}\right) - I_0\left( \frac{ar}{\varepsilon^2}\right)  \right] & r \geq a
 \end{cases}
\end{equation}

\begin{figure}[ht]
\centering
    \includegraphics[width=0.45\linewidth]{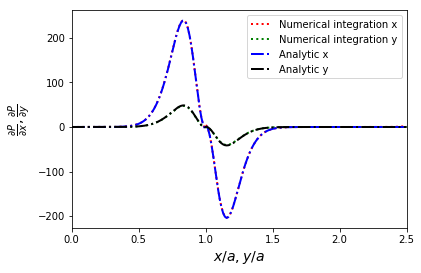}
        \includegraphics[width=0.45\linewidth]{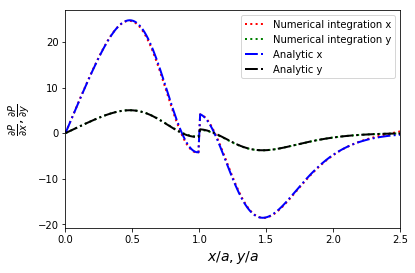}
            \includegraphics[width=0.45\linewidth]{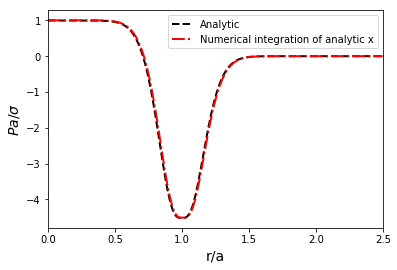}
        \includegraphics[width=0.45\linewidth]{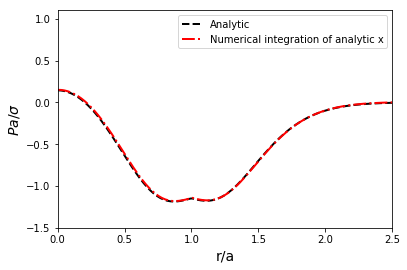}
    \caption{Comparison of numerical integration and the exact solution with $a/\varepsilon = 6$ (left) and $a/\varepsilon = 2$ (right). Note the discontinuity the forms when $a/\varepsilon  = 2$. This corresponds to the negative pressures seen in the simulations. In this case the pressure (bottom row) does not become negative, but it is significantly reduced from its original value. The values are taken along the line $(x, y) = (r\cos(0.2), r\sin(0.2)).$}
    \label{fig:2dplot}
\end{figure}

The solution from eqs. \ref{eq:dpdx},  \ref{eq:dpdy} and \ref{eq:2dpressure} are plotted in fig. \ref{fig:2dplot}, with a comparison to numerically integrating eq. \ref{eq:2dequation}. The radius of support for which the gradient of the pressure is non-zero corresponds to a region of $3.5 \varepsilon$ around the interface. The pressure profiles vary continuously across the interface, instead of a sharp discontinuity, and the behavior is quantitatively similar to that of MD simulations \citep{Masuda2011, Nakamura2011}. For $a/\varepsilon =2$, the pressure difference between the inside and outside of the bubble, $P_{\varepsilon, in}-P_{\varepsilon, out}$ is less than $\sigma/a$, showing that the pressure difference deviates from the Young-Laplace law in this case.  The pressure difference begins to deviate from the Young-Laplace law at $a/\varepsilon \approx 3.5$, and $P_{\varepsilon, in}-P_{\varepsilon, out}$ decreases as $a/\varepsilon$ decreases. 

\section{Three-dimensional bubble}

We consider a sphere with radius $a$ centered at the origin. Let the point where we desire to calculate the pressure be $\bx = (x_i, y_i, z_i)$. For ease, of computation, we switch to spherical coordinates, so $\bx = (r_i \sin \theta_i \cos \varphi_i, r_i \sin \theta_i \sin \varphi_i,  r_i \cos \theta_i )$. Denote the point $\by$ as $(r \sin \theta \cos \varphi,  r \sin \theta \sin \varphi,  r \cos \theta ).$ Note that $|\bx - \by|^2 = r^2 + r_i^2 -2 r r_i \sin \theta\sin \theta_i\cos(\varphi - \varphi_i)-2rr_i\cos \theta \cos \theta_i$. Also, $s(\bx, \by)$ will depend only on $r$ and $r_i$. 

We will first consider the derivative with respect to $\varphi$:
\begin{align}
\frac{1}{r_i\sin\theta_i}\frac{\partial P}{\partial \varphi_i} &= - \int_{0}^\infty \int_0^\pi \int_0^{2\pi}s(r,r_i) e^{-\frac{r^2 + r_i^2 -2 r r_i \sin \theta\sin \theta_i\cos(\varphi - \varphi_i)-2rr_i\cos \theta \cos \theta_i}{2 \varepsilon^2}}r^3\sin(\varphi - \varphi_i)\sin^2\theta \; d\varphi d\theta dr  \nonumber \\
& \quad +A \int_{0}^\infty \int_0^\pi \int_0^{2\pi}s(r,r_i) e^{-\frac{r^2 + r_i^2 -2 r r_i \sin \theta\sin \theta_i\cos(\varphi - \varphi_i)-2rr_i\cos \theta \cos \theta_i}{2 \varepsilon_0^2}}r^3\sin(\varphi - \varphi_i)\sin^2\theta \; d\varphi d\theta dr\nonumber \\
& =  0.
\end{align}

Now, consider the derivative with respect to $\theta$: 
\begin{align}
\frac{1}{r_i}\frac{\partial P}{\partial \theta_i} &=  -\int_{0}^\infty \int_0^\pi \int_0^{2\pi} s(r,r_i) e^{-\frac{r^2 + r_i^2 -2 r r_i \sin \theta\sin \theta_i\cos(\varphi - \varphi_i)-2rr_i\cos \theta \cos \theta_i}{2 \varepsilon^2}} r^3 [\sin^2\theta\cos\theta_i\cos(\varphi - \varphi_i) \nonumber \\
&\quad\quad  -\sin\theta \cos \theta \sin \theta_i] \; d\varphi d\theta dr  \nonumber \\
&\quad +A \int_{0}^\infty \int_0^\pi \int_0^{2\pi} s(r,r_i) e^{-\frac{r^2 + r_i^2 -2 r r_i \sin \theta\sin \theta_i\cos(\varphi - \varphi_i)-2rr_i\cos \theta \cos \theta_i}{2 \varepsilon_0^2}} r^3 [\sin^2\theta\cos\theta_i\cos(\varphi - \varphi_i) \nonumber \\
&\quad\quad  -\sin\theta \cos \theta \sin \theta_i] \; d\varphi d\theta dr   \label{eq:six}.
\end{align}
Let $\kappa = 2\varepsilon^2$ or $2\varepsilon_0^2$ and consider one term of the integral in eq. \ref{eq:six}:
\begin{align}
&\int_{0}^\infty \int_0^\pi \int_0^{2\pi} s(r,r_i) e^{-\frac{r^2 + r_i^2 -2 r r_i \sin \theta\sin \theta_i\cos(\varphi - \varphi_i)-2rr_i\cos \theta \cos \theta_i}{\kappa}} r^3 [\sin^2\theta\cos\theta_i\cos(\varphi - \varphi_i) \nonumber \\
&\quad\quad  -\sin\theta \cos \theta \sin \theta_i] \; d\varphi d\theta dr  \nonumber \\
&=  2\pi \int_{0}^\infty s(r,r_i) r^3 e^{-\frac{r^2 + r_i^2 }{\kappa}}  \int_0^\pi e^{\frac{ 2rr_i\cos \theta \cos \theta_i}{\kappa}}  \left[\sin^2\theta\cos\theta_iI_1\left(\frac{2 r r_i \sin \theta\sin \theta_i}{\kappa}\right) \right. \nonumber\\
&\quad  \left. -\sin\theta \cos \theta \sin \theta_iI_0\left(\frac{ 2 r r_i \sin \theta\sin \theta_i}{\kappa}\right)\right] \; d\theta dr \label{dpdr_partial} 
\end{align}
Consider integration by parts on the last term in eq. \ref{dpdr_partial} with $u = e^{\frac{ 2rr_i\cos \theta \cos \theta_i}{\kappa}} $ and \\ \noindent $dv = \sin\theta \cos \theta \sin \theta_iI_0\left(\frac{ 2 r r_i \sin \theta\sin \theta_i}{\kappa}\right) \; d\theta$. For ease of notation let $C = \frac{2r r_i}{\kappa}.$ Then, 
$\int dv = \frac{\sin\theta I_1\left(C \sin \theta\sin \theta_i \right)}{C}  $ and 
$du = -C\cos\theta_i\sin\theta e^{C \cos \theta \cos \theta_i} d\theta$, 
so 
\begin{align}
&\int_0^\pi e^{C \cos \theta \cos \theta_i }  \sin\theta \cos \theta \sin \theta_iI_0\left(C \sin \theta\sin \theta_i\right) \; d\theta  \nonumber \\
& = \left[e^{C \cos \theta \cos \theta_i } \frac{\sin\theta I_1\left(C \sin \theta\sin \theta_i \right)}{C} \right]_0^\pi + \int_0^\pi C\cos\theta_i\sin\theta e^{C \cos \theta \cos \theta_i} \frac{\sin\theta I_1\left(C \sin \theta\sin \theta_i \right)}{C} \; d\theta \nonumber \\
& = \int_0^\pi \sin^2\theta \cos\theta_i e^{C \cos \theta \cos \theta_i} I_1\left(C \sin \theta\sin \theta_i \right) \; d\theta \nonumber
\end{align}
Note that this is exactly the quantity in the first part of the integral in eq. \ref{dpdr_partial} but will have opposite sign, so therefore
\begin{equation}
\frac{1}{r_i}\frac{\partial P}{\partial \theta_i} = 0.
\end{equation}

Now we turn to the derivative with respect to $r$:
\begin{align}
\frac{\partial P}{\partial r_i} &= - \int_{0}^\infty \int_0^\pi \int_0^{2\pi} s(r,r_i) e^{-\frac{r^2 + r_i^2 -2 r r_i \sin \theta\sin \theta_i\cos(\varphi - \varphi_i)-2rr_i\cos \theta \cos \theta_i}{2\varepsilon^2}} r^2\sin\theta [r\cos\theta \cos \theta_i \nonumber  +r\sin\theta \sin \theta_i \cos(\varphi-\varphi_i) \\
 &\quad\quad - r_i] \; d\varphi d\theta dr  \nonumber \\
&\quad +A  \int_{0}^\infty \int_0^\pi \int_0^{2\pi} s(r,r_i) e^{-\frac{r^2 + r_i^2 -2 r r_i \sin \theta\sin \theta_i\cos(\varphi - \varphi_i)-2rr_i\cos \theta \cos \theta_i}{2\varepsilon_0^2}} r^2\sin\theta [r\cos\theta \cos \theta_i \nonumber  +r\sin\theta \sin \theta_i \cos(\varphi-\varphi_i) \\
 &\quad\quad - r_i] \; d\varphi d\theta dr \label{eq:dpdr3d}
 \end{align}
 Let $\kappa$ be defined as above and consider one of the integrals in eq. \ref{eq:dpdr3d}.
\begin{align}
 &  \int_{0}^\infty \int_0^\pi \int_0^{2\pi} s(r,r_i) e^{-\frac{r^2 + r_i^2 -2 r r_i \sin \theta\sin \theta_i\cos(\varphi - \varphi_i)-2rr_i\cos \theta \cos \theta_i}{\kappa}} r^2\sin\theta [r\cos\theta \cos \theta_i \nonumber  +r\sin\theta \sin \theta_i \cos(\varphi-\varphi_i) - r_i] \; d\varphi d\theta dr  \nonumber \\
&=  2\pi \int_{0}^\infty s(r,r_i) e^{-\frac{r^2 + r_i^2}{\kappa}} r^2 \int_0^\pi e^{\frac{2rr_i\cos \theta \cos \theta_i}{\kappa}}  
\left[ 
\left(r\cos\theta \sin\theta  \cos \theta_i  - r_i \sin\theta  \right) I_0\left(\frac{ 2 r r_i \sin \theta\sin \theta_i}{\kappa}\right) \right.\nonumber \\
&\quad +\left. \left(r\sin^2\theta \sin \theta_i  \right) I_1\left(\frac{ 2 r r_i \sin \theta\sin \theta_i}{\kappa}\right) \right]\;  d\theta dr  
\end{align}
Again let $C = \frac{2 rr_i}{\kappa}. $
We assume $\theta_i = 0$. Then, 
\begin{align}
&2\pi \int_{0}^\infty s(r,r_i) e^{-\frac{r^2 + r_i^2}{\kappa}} r^2 \int_0^\pi e^{C\cos \theta}  
\left[ \left(r\cos\theta \sin\theta - r_i \sin\theta  \right) \right]\;  d\theta dr  \nonumber \\
&=2\pi \int_{0}^\infty s(r,r_i) e^{-\frac{r^2 + r_i^2}{\kappa}} r^2   
\left[r \frac{e^{C\cos\theta}(1-C\cos\theta)}{C^2} +r_i\frac{e^{C\cos\theta}}{C}   \right]_0^\pi \;  dr \nonumber \\
&=-2\pi \int_{0}^\infty s(r,r_i) e^{-\frac{r^2 + r_i^2}{\kappa}}    
\left[\left(\frac{\kappa^2}{2r_i^2}+ \kappa  \right) r\sinh\left( \frac{2 rr_i}{\kappa}\right)-\frac{\kappa }{r_i}r^2\cosh\left( \frac{2 rr_i}{\kappa}\right) \right] \;   dr \nonumber 
\end{align}
Note that $s(r, r_i)$ will be constant for $r \in [0, a]$ and $r \in (a, \infty)$, so 
\begin{align}
&-2\pi \int_{0}^\infty s(r,r_i) e^{-\frac{r^2 + r_i^2}{\kappa}}    
\left[\left(\frac{\kappa^2}{2r_i^2}+ \kappa  \right) r\sinh\left( \frac{2 rr_i}{\kappa}\right)-\frac{\kappa }{r_i}r^2\cosh\left( \frac{2 rr_i}{\kappa}\right) \right] \;   dr \nonumber  \\
&\quad=-2\pi \int_{0}^a s_1 e ^{-\frac{r^2 + r_i^2}{\kappa}}    
\left[\left(\frac{\kappa^2}{2r_i^2}+ \kappa  \right) r\sinh\left( \frac{2 rr_i}{\kappa}\right)-\frac{\kappa }{r_i}r^2\cosh\left( \frac{2 rr_i}{\kappa}\right)\right] \;   dr \nonumber \\
&\quad\quad-2\pi \int_{a}^\infty s_2 e^{-\frac{r^2 + r_i^2}{\kappa}}    
\left[\left(\frac{\kappa^2}{2r_i^2}+ \kappa  \right) r\sinh\left( \frac{2 rr_i}{\kappa}\right)-\frac{\kappa }{r_i}r^2\cosh\left( \frac{2 rr_i}{\kappa}\right) \right] \;   dr  \label{eq:four}
\end{align}
where $s_1$ and $s_2$ are constants that depend on whether $r_i \in [0, a]$ or $r_i \in (a, \infty)$.

We break the integral in eq. \ref{eq:four} into two parts: 
\begin{align}
J_0 &:=-\frac{2 \pi \kappa}{r_i} e^{-\frac{ r_i^2}{\kappa}}   \int  s e^{-\frac{r^2 }{\kappa}}    r^2 \cosh\left( \frac{2rr_i}{\kappa}\right) dr \nonumber \\
  &= -\frac{ \pi \kappa^2}{8r_i} \left[\frac{\sqrt{\kappa \pi}(4r_i^2+2\kappa)}{\kappa} \left( \erf\left(\frac{r-r_i}{\sqrt{\kappa}}\right)+ \erf\left(\frac{r+r_i}{\sqrt{\kappa}}\right) 
 \right) -4 e^{-\frac{(r_i+r)^2}{\kappa}} \left( (r_i+r)e^{\frac{4r_ir}{\kappa}} -r_i + r \right) \right] \nonumber 
\end{align}
We need to evaluate $J_0$ at $0$, $a$, and $\infty$. At zero, we have: 
\begin{align}
J_0(0) &= -\frac{ \pi \kappa^2}{8r_i} \left[\frac{\sqrt{\kappa \pi}(4r_i^2+2\kappa)}{\kappa} \left( \erf\left(\frac{-r_i}{\sqrt{\kappa}}\right)+ \erf\left(\frac{r_i}{\sqrt{\kappa}}\right) 
 \right) \right.  \left.  -4 e^{-\frac{r_i^2}{\kappa}} \left( r_i -r_i  \right) \right] = 0 \nonumber 
\end{align}
At $a$ we have: 
\begin{align}
  J_0(a)&= -\frac{ \pi \kappa^2}{8r_i} \left[\frac{\sqrt{\kappa \pi}(4r_i^2+2\kappa)}{\kappa} \left[ \erf\left(\frac{a-r_i}{\sqrt{\kappa}}\right)+ \erf\left(\frac{a+r_i}{\sqrt{\kappa}}\right) 
 \right] -4 e^{-\frac{(r_i+r)^2}{\kappa}} \left( (r_i+a)e^{\frac{4r_ia}{\kappa}} -r_i + a \right) \right] \nonumber \\
 &=-\frac{( \pi \kappa)^{3/2}(2r_i^2+\kappa)}{4r_i}  \left[ \erf\left(\frac{a-r_i}{\sqrt{\kappa}}\right)+ \erf\left(\frac{a+r_i}{\sqrt{\kappa}}\right) 
 \right] +\frac{ \pi \kappa^2}{2r_i} e^{-\frac{(r_i+r)^2}{\kappa}} \left( (r_i+a)e^{\frac{4r_ia}{\kappa}} -r_i + a \right) \nonumber 
\end{align}
And finally, 
\begin{align}
\lim_{r\rightarrow \infty} J_0(r) &=-\lim_{r\rightarrow \infty}  \frac{ \pi \kappa^2}{8r_i} \left[\frac{\sqrt{\kappa \pi}(4r_i^2+2\kappa)}{\kappa} \left( \erf\left(\frac{r-r_i}{\sqrt{\kappa}}\right)+ \erf\left(\frac{r+r_i}{\sqrt{\kappa}}\right) 
 \right) -4 e^{-\frac{(r_i+r)^2}{\kappa}} \left( (r_i+r)e^{\frac{4r_ir}{\kappa}} -r_i + r \right) \right] \nonumber \\
 &=- \frac{ \pi \kappa^2}{8r_i} \frac{\sqrt{\kappa \pi}(4r_i^2+2\kappa)}{\kappa} \left(1+1 \right)  + \lim_{r\rightarrow \infty}  \frac{ \pi \kappa^2}{8r_i} \left[4 e^{-\frac{(r_i+r)^2}{\kappa}} \left( (r_i+r)e^{\frac{4r_ir}{\kappa}} -r_i + r \right) \right] \nonumber \\
  &= -\frac{ (\pi \kappa)^{3/2}(4r_i^2+2\kappa)}{4r_i}  \nonumber 
\end{align}

\noindent The second part of the integral in eq. \ref{eq:four} gives:
\begin{align}
J_1 &:= 2\pi \left( \frac{\kappa^2}{2r_i^2}+ \kappa \right) e^{-\frac{ r_i^2}{\kappa}}   \int  e^{-\frac{r^2 }{\kappa}}    
r \sinh\left( \frac{2rr_i}{\kappa}\right) \;   dr  \nonumber \\
&= \frac{(\pi \kappa )^{3/2} (\kappa + 2r_i^2)}{4r_i}\left[ \erf\left(\frac{r-r_i}{\sqrt{\kappa}}\right)+ \erf\left(\frac{r+r_i}{\sqrt{\kappa}} \right) \right] - \frac{\pi \kappa }{2} \left( \frac{\kappa^2}{2r_i^2}+ \kappa \right)  e^{-\frac{(r_i + r)^2}{\kappa}}\left[e^{\frac{4r_i r}{\kappa}} - 1\right] \nonumber 
\end{align}
We again evaluate $J_1$ at $0, a$, and find the limit as $r$ approaches $\infty$. 
\begin{align}
J_1(0) &= \frac{(\pi \kappa )^{3/2} (\kappa + 2r_i^2)}{4r_i}\left[ \erf\left(\frac{-r_i}{\sqrt{\kappa}}\right)+ \erf\left(\frac{r_i}{\sqrt{\kappa}} \right) \right] - \frac{\pi \kappa }{2} \left( \frac{\kappa^2}{2r_i^2}+ \kappa \right)  e^{-\frac{(r_i )^2}{\kappa}}\left[e^{0} - 1\right]  = 0\nonumber 
\end{align}

\begin{align}
J_1(a) &=  \frac{(\pi \kappa )^{3/2} (\kappa + 2r_i^2)}{4r_i}\left[ \erf\left(\frac{a-r_i}{\sqrt{\kappa}}\right)+ \erf\left(\frac{a+r_i}{\sqrt{\kappa}} \right) \right] - \frac{\pi \kappa }{2} \left( \frac{\kappa^2}{2r_i^2}+ \kappa \right)  e^{-\frac{(r_i + a)^2}{\kappa}}\left[e^{\frac{4r_i a}{\kappa}} - 1\right] \nonumber 
\end{align}

\begin{align}
\lim_{r\rightarrow \infty}  J_1(r) &=  \lim_{r \rightarrow \infty} \frac{(\pi \kappa )^{3/2} (\kappa + 2r_i^2)}{4r_i}\left[ \erf\left(\frac{r-r_i}{\sqrt{\kappa}}\right)+ \erf\left(\frac{r+r_i}{\sqrt{\kappa}} \right) \right] - \frac{\pi \kappa }{2} \left( \frac{\kappa^2}{2r_i^2}+ \kappa \right)  e^{-\frac{(r_i + r)^2}{\kappa}}\left[e^{\frac{4r_i r}{\kappa}} - 1\right] \nonumber  \\
&=  \frac{(\pi \kappa )^{3/2} (\kappa + 2r_i^2)}{2r_i} \nonumber 
\end{align}

\noindent Note that $J_0(0) + J_1(0) = 0$ and $\lim_{r\rightarrow \infty} J_0(r) + J_1(r) = -\frac{ (\pi \kappa)^{3/2}(\kappa + 2r_i^2)}{2r_i} +\frac{(\pi \kappa )^{3/2} (\kappa + 2r_i^2)}{2r_i}$.
Therefore, 
\begin{align*}
2\pi  e^{-\frac{ r_i^2}{\kappa}}   \int_{0}^a  e^{-\frac{r^2 }{\kappa}}    
\left[\left(\frac{\kappa r^2}{r_i}-\kappa r \right)\cosh\left( \frac{2rr_i}{\kappa}\right)-\frac{\kappa^2 r}{2r_i^2} \sinh\left( \frac{2rr_i}{\kappa}\right) \right] \;   dr &=J_0(a)-J_0(0)+J_1(a)-J_1(0)  \nonumber \\
&=\frac{ \pi \kappa^2}{2r_i} e^{-\frac{(r_i+a)^2}{\kappa}} \left[  \frac{2r_ia-\kappa}{2r_i}e^{\frac{4r_ia}{\kappa}} +  \frac{2r_ia+\kappa}{2r_i } \right]
\end{align*}

\noindent Similarly, 
\begin{align*}
2\pi  e^{-\frac{ r_i^2}{\kappa}}   \int_{a}^\infty  e^{-\frac{r^2 }{\kappa}}    
\left[\left(\frac{\kappa r^2}{r_i}-\kappa r \right)\cosh\left( \frac{2rr_i}{\kappa}\right)-\frac{\kappa^2 r}{2r_i^2} \sinh\left( \frac{2rr_i}{\kappa}\right) \right] \;   dr &=\lim_{r\rightarrow \infty} J_0(r)  - J_0(a)+\lim_{r\rightarrow \infty} J_1(r) -J_1(a) \nonumber \\
&=-\frac{ \pi \kappa^2}{2r_i} e^{-\frac{(r_i+a)^2}{\kappa}} \left[  \frac{2r_ia-\kappa}{2r_i}e^{\frac{4r_ia}{\kappa}} +  \frac{2r_ia+\kappa}{2r_i } \right]
\end{align*}

\noindent Thus, we have, for $r \in [0, a]$ and $\theta_i = 0$:    
\boxalign{ \begin{align}
 \frac{\partial P}{\partial \theta_i} &= 0 \\
  \frac{\partial P}{\partial \varphi_i} &= 0 \\
 \frac{\partial P}{\partial r_i} &=  \quad\;\;-(1-2H[r_i-a])(s_{\alpha \alpha}-s_{\alpha \beta})\frac{ 2 \pi \varepsilon^4}{r_i} e^{-\frac{(r_i+a)^2}{2\varepsilon^2}} \left[  \frac{r_ia-\varepsilon^2}{r_i}e^{\frac{4r_ia}{2\varepsilon^2}} +  \frac{r_ia+\varepsilon^2}{r_i } \right] \nonumber \\
  &\quad +A (1-2H[r_i-a])(s_{\alpha \alpha}-s_{\alpha \beta})\frac{ 2 \pi \varepsilon_0^4}{r_i} e^{-\frac{(r_i+a)^2}{2\varepsilon_0^2}} \left[  \frac{r_ia-\varepsilon_0^2}{r_i}e^{\frac{4r_ia}{2\varepsilon_0^2}} +  \frac{r_ia+\varepsilon_0^2}{r_i } \right] \label{eq:dpdrfinal}
 \end{align}}
 
  Note that since the partial derivatives with respect to $\theta_i$ and $\varphi_i$ are zero, 
 $  \frac{\partial P}{\partial x} = \frac{x}{|\bx|} \frac{\partial P}{\partial r_i}$,  $  \frac{\partial P}{\partial y} = \frac{y}{|\bx|} \frac{\partial P}{\partial r_i}$, and  $  \frac{\partial P}{\partial z} = \frac{z}{|\bx|} \frac{\partial P}{\partial r_i}$.

To find the pressure, we need to integrate equation \ref{eq:dpdrfinal}:
\begin{align}
P(r) - P(\infty) &= -\int_\infty^r (1-2H[r_i-a])(s_{\alpha \alpha}-s_{\alpha \beta})\frac{ 2 \pi \varepsilon^4}{r_i} e^{-\frac{(r_i+a)^2}{2\varepsilon^2}} \left[  \frac{r_ia-\varepsilon^2}{r_i}e^{\frac{4r_ia}{2\varepsilon^2}} +  \frac{r_ia+\varepsilon^2}{r_i } \right] \; dr_i \nonumber \\
&+A \int_\infty^r (1-2H[r_i-a])(s_{\alpha \alpha}-s_{\alpha \beta})\frac{ 2 \pi \varepsilon_0^4}{r_i} e^{-\frac{(r_i+a)^2}{2\varepsilon_0^2}} \left[  \frac{r_ia-\varepsilon_0^2}{r_i}e^{\frac{4r_ia}{2\varepsilon_0^2}} +  \frac{r_ia+\varepsilon_0^2}{r_i } \right] \; dr_i \label{eq:3dpint}
\end{align}
Note that eq. \ref{eq:3dpint} is equivalent to 
\begin{align}
P(r) - P(\infty) &= -(s_{\alpha \alpha}-s_{\alpha \beta}) 4 \pi \varepsilon^4 e^{-\frac{a^2}{2\varepsilon^2}} \int_\infty^r (1-2H[r_i-a]) e^{-\frac{r_i^2}{2\varepsilon^2}} \left[ \frac{a}{r_i} \cosh\left( \frac{r_i a}{\varepsilon^2}\right)-\frac{\varepsilon^2}{r_i^2} \sinh\left( \frac{r_i a}{\varepsilon^2}\right)\right] \; dr_i  \nonumber \\
&+A(s_{\alpha \alpha}-s_{\alpha \beta}) 4 \pi \varepsilon_0^4 e^{-\frac{a^2}{2\varepsilon_0^2}} \int_\infty^r (1-2H[r_i-a]) e^{-\frac{r_i^2}{2\varepsilon_0^2}} \left[ \frac{a}{r_i} \cosh\left( \frac{r_i a}{\varepsilon_0^2}\right)-\frac{\varepsilon_0^2}{r_i^2} \sinh\left( \frac{r_i a}{\varepsilon_0^2}\right)\right] \; dr_i  \label{eq:eight}
\end{align}
  
\noindent  For $r> a$,  the first integral in eq. \ref{eq:eight} reduces to
 \begin{equation}
J_2(r, \varepsilon): = (s_{\alpha \alpha}-s_{\alpha \beta}) 4 \pi \varepsilon^4 e^{-\frac{a^2}{2\varepsilon^2}} \int_\infty^r e^{-\frac{r_i^2}{2\varepsilon^2}} \left[ \frac{a}{r_i} \cosh\left( \frac{r_i a}{\varepsilon^2}\right)-\frac{\varepsilon^2}{r_i^2} \sinh\left( \frac{r_i a}{\varepsilon^2}\right)\right] \; dr_i  \label{eq:nine}
\end{equation}
For $a > r $, we can then consider 
   \begin{equation}
J_1(r, \varepsilon) :=-J_2(a, \varepsilon) -(s_{\alpha \alpha}-s_{\alpha \beta}) 4 \pi \varepsilon^4 e^{-\frac{a^2}{2\varepsilon^2}} \int_a^r e^{-\frac{r_i^2}{2\varepsilon^2}} \left[ \frac{a}{r_i} \cosh\left( \frac{r_i a}{\varepsilon^2}\right)-\frac{\varepsilon^2}{r_i^2} \sinh\left( \frac{r_i a}{\varepsilon^2}\right)\right] \; dr_i 
\end{equation}
  
\noindent Consider integration by parts on the first term in eq. \ref{eq:nine}. 
\begin{align}
& \int_\infty^r e^{-\frac{r_i^2}{2\varepsilon^2}}  \frac{a}{r_i} \cosh\left( \frac{r_i a}{\varepsilon^2}\right)\nonumber \\
  &=\left[ \frac{\varepsilon^2}{r_i} e^{-\frac{r_i^2}{2\varepsilon^2}} \sinh\left( \frac{r_i a}{\varepsilon^2}\right)  \right]_\infty^r + \int_\infty^r e^{-\frac{r_i^2}{2\varepsilon^2}} \sinh\left( \frac{r_i a}{\varepsilon^2}\right) \left(1 + \frac{\varepsilon^2}{r_i^2}\right) \; dr_i\nonumber \\
 &= \frac{\varepsilon^2}{r} e^{-\frac{r^2}{2\varepsilon^2}} \sinh\left( \frac{r a}{\varepsilon^2}\right) -\frac{\sqrt{2\pi}\varepsilon}{4}e^{\frac{a^2}{2\varepsilon^2}} \left[ \erf\left( \frac{a-r_i}{\sqrt{2}\varepsilon}\right)+ \erf\left( \frac{a+r_i}{\sqrt{2}\varepsilon}\right) \right]_\infty^r 
  + \int_\infty^r e^{-\frac{r_i^2}{2\varepsilon^2}} \sinh\left( \frac{r_i a}{\varepsilon^2}\right) \frac{\varepsilon^2}{r_i^2}\; dr_i \nonumber \\
   &= \frac{\varepsilon^2}{r} e^{-\frac{r^2}{2\varepsilon^2}} \sinh\left( \frac{r a}{\varepsilon^2}\right) -\frac{\sqrt{2\pi}\varepsilon}{4}e^{\frac{a^2}{2\varepsilon^2}} \left[ \erf\left( \frac{a-r}{\sqrt{2}\varepsilon}\right)+ \erf\left( \frac{a+r}{\sqrt{2}\varepsilon}\right) \right]
  + \int_\infty^r e^{-\frac{r_i^2}{2\varepsilon^2}} \sinh\left( \frac{r_i a}{\varepsilon^2}\right) \frac{\varepsilon^2}{r_i^2}\; dr_i \label{eq:ten}.
 \end{align}
 
\noindent Combining eq. \ref{eq:ten} with eq. \ref{eq:nine} gives: 
 \begin{align}
 J_2(r, \varepsilon): &= (s_{\alpha \alpha}-s_{\alpha \beta}) 4 \pi \varepsilon^4 e^{-\frac{a^2}{2\varepsilon^2}} \left(\frac{\varepsilon^2}{r} e^{-\frac{r^2}{2\varepsilon^2}} \sinh\left( \frac{r a}{\varepsilon^2}\right) -\frac{\sqrt{2\pi}\varepsilon}{4}e^{\frac{a^2}{2\varepsilon^2}} \left[ \erf\left( \frac{a-r}{\sqrt{2}\varepsilon}\right)+ \erf\left( \frac{a+r}{\sqrt{2}\varepsilon}\right) \right]
  \right. \nonumber \\
  &\left.\quad+ \int_\infty^r e^{-\frac{r_i^2}{2\varepsilon^2}} \sinh\left( \frac{r_i a}{\varepsilon^2}\right) \frac{\varepsilon^2}{r_i^2}\; dr_i- \int_\infty^r e^{-\frac{r_i^2}{2\varepsilon^2}} \frac{\varepsilon^2}{r_i^2} \sinh\left( \frac{r_i a}{\varepsilon^2}\right) \; dr_i  \right) \nonumber \\
   &= (s_{\alpha \alpha}-s_{\alpha \beta}) 4 \pi \varepsilon^4  \left(\frac{\varepsilon^2}{r} e^{-\frac{a^2 + r^2}{2\varepsilon^2}} \sinh\left( \frac{r a}{\varepsilon^2}\right) -\frac{\sqrt{2\pi}\varepsilon}{4} \left[ \erf\left( \frac{a-r}{\sqrt{2}\varepsilon}\right)+ \erf\left( \frac{a+r}{\sqrt{2}\varepsilon}\right) \right]
  \right) \nonumber 
 \end{align}
 
\noindent Note that $J_2(a, \varepsilon) = (s_{\alpha \alpha}-s_{\alpha \beta}) 4 \pi \varepsilon^4  \left(\frac{\varepsilon^2}{r} e^{-\frac{a^2}{\varepsilon^2}} \sinh\left( \frac{a^2}{\varepsilon^2}\right) -\frac{\sqrt{2\pi}\varepsilon}{4} \left[ \erf\left( \frac{2a}{\sqrt{2}\varepsilon}\right) \right]
  \right)$. Therefore, 
     \begin{align}
J_1(r, \varepsilon) &=(s_{\alpha \alpha}-s_{\alpha \beta}) 4 \pi \varepsilon^4  \left(\frac{\varepsilon^2}{a} e^{-\frac{a^2}{\varepsilon^2}} \sinh\left( \frac{a^2}{\varepsilon^2}\right) -\frac{\sqrt{2\pi}\varepsilon}{4} \left[ \erf\left( \frac{2a}{\sqrt{2}\varepsilon}\right) \right]
  \right) \nonumber \\
  &\quad -(s_{\alpha \alpha}-s_{\alpha \beta}) 4 \pi \varepsilon^4 e^{-\frac{a^2}{2\varepsilon^2}} \int_a^r e^{-\frac{r_i^2}{2\varepsilon^2}} \left[ \frac{a}{r_i} \cosh\left( \frac{r_i a}{\varepsilon^2}\right)-\frac{\varepsilon^2}{r_i^2} \sinh\left( \frac{r_i a}{\varepsilon^2}\right)\right] \; dr_i   \nonumber \\
  &=(s_{\alpha \alpha}-s_{\alpha \beta}) 4 \pi \varepsilon^4  \left(\frac{\varepsilon^2}{r} e^{-\frac{a^2}{\varepsilon^2}} \sinh\left( \frac{a^2}{\varepsilon^2}\right) -\frac{\sqrt{2\pi}\varepsilon}{4} \left[ \erf\left( \frac{2a}{\sqrt{2}\varepsilon}\right) \right]
  \right)  \nonumber \\
  &\quad -(s_{\alpha \alpha}-s_{\alpha \beta}) 4 \pi \varepsilon^4 \left\{\left[ \frac{\varepsilon^2}{r_i} e^{-\frac{a^2 + r_i^2}{2\varepsilon^2}} \sinh\left( \frac{r_i a}{\varepsilon^2}\right)  \right]_a^r -\frac{\sqrt{2\pi}\varepsilon}{4} \left[ \erf\left( \frac{a-r_i}{\sqrt{2}\varepsilon}\right)+ \erf\left( \frac{a+r_i}{\sqrt{2}\varepsilon}\right) \right]_a^r \right\}  \nonumber  \\
      &=-(s_{\alpha \alpha}-s_{\alpha \beta}) 4 \pi \varepsilon^4  \left\{ -2 \frac{\varepsilon^2}{a} e^{-\frac{a^2}{\varepsilon^2}} \sinh\left( \frac{a^2}{\varepsilon^2}\right) +\frac{\sqrt{2\pi}\varepsilon}{2} \erf\left( \frac{2a}{\sqrt{2}\varepsilon}\right) +  \frac{\varepsilon^2}{r} e^{-\frac{a^2 + r^2}{2\varepsilon^2}} \sinh\left( \frac{r a}{\varepsilon^2}\right)     \right. \nonumber \\
  &\quad \left.-\frac{\sqrt{2\pi}\varepsilon}{4} \left[ \erf\left( \frac{a-r}{\sqrt{2}\varepsilon}\right)+ \erf\left( \frac{a+r}{\sqrt{2}\varepsilon}\right) \right] \right\}  
\end{align}

\begin{figure}[ht]
\centering
    \includegraphics[width=0.45\linewidth]{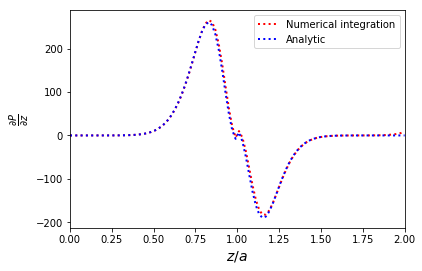}
        \includegraphics[width=0.45\linewidth]{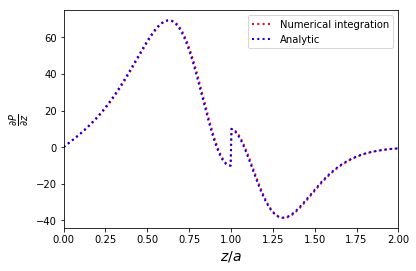}
        \includegraphics[width=0.45\linewidth]{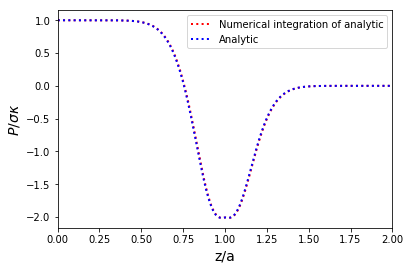}
                \includegraphics[width=0.45\linewidth]{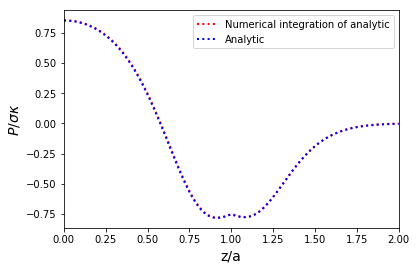}
    \caption{(Top) Comparison of numerical integration of eq. \ref{eq:dpdr3d} and the analytic solution in eq. \ref{eq:dpdrfinal} with $a/\varepsilon = 6$, $ a =1$ (left) and  $a/\varepsilon = 3$, $ a =1$ (right). (Bottom) Comparison of the analytic solution in eq. \ref{eq:3dpressure} and numerically integrating eq.  \ref{eq:dpdrfinal}.}
    \label{fig:3dplot}
\end{figure}

\noindent So, we get: 
\boxalign{
\begin{align}
P(r) - P(\infty) = \begin{cases}
(s_{\alpha \alpha}-s_{\alpha \beta}) 4 \pi \varepsilon^5  \left(\frac{\varepsilon}{r} e^{-\frac{a^2 + r^2}{2\varepsilon^2}} \sinh\left( \frac{r a}{\varepsilon^2}\right) -\frac{\sqrt{2\pi}}{4} \left[ \erf\left( \frac{a-r}{\sqrt{2}\varepsilon}\right)+ \erf\left( \frac{a+r}{\sqrt{2}\varepsilon}\right) \right]
  \right) \\
  \quad\quad   -A(s_{\alpha \alpha}-s_{\alpha \beta}) 4 \pi \varepsilon_0^5  \left(\frac{\varepsilon_0}{r} e^{-\frac{a^2 + r^2}{2\varepsilon_0^2}} \sinh\left( \frac{r a}{\varepsilon_0^2}\right) -\frac{\sqrt{2\pi}}{4} \left[ \erf\left( \frac{a-r}{\sqrt{2}\varepsilon_0}\right)+ \erf\left( \frac{a+r}{\sqrt{2}\varepsilon_0}\right) \right]
  \right) & r \geq a\\
-(s_{\alpha \alpha}-s_{\alpha \beta}) 4 \pi \varepsilon^5  \left\{ -2 \frac{\varepsilon}{a} e^{-\frac{a^2}{\varepsilon^2}} \sinh\left( \frac{a^2}{\varepsilon^2}\right) +\frac{\sqrt{2\pi}}{2} \erf\left( \frac{2a}{\sqrt{2}\varepsilon}\right) +\frac{\varepsilon}{r} e^{-\frac{a^2 + r^2}{2\varepsilon^2}} \sinh\left( \frac{r a}{\varepsilon^2}\right)   \right. \\
  \quad\quad  \left.  -\frac{\sqrt{2\pi}}{4} \left[ \erf\left( \frac{a-r}{\sqrt{2}\varepsilon}\right)+ \erf\left( \frac{a+r}{\sqrt{2}\varepsilon}\right) \right] \right\}  +A(s_{\alpha \alpha}-s_{\alpha \beta}) 4 \pi \varepsilon_0^5  \left\{ -2 \frac{\varepsilon_0}{a} e^{-\frac{a^2}{\varepsilon_0^2}} \sinh\left( \frac{a^2}{\varepsilon_0^2}\right) \right. \\
   \quad\quad   \left. +\frac{\sqrt{2\pi}}{2} \erf\left( \frac{2a}{\sqrt{2}\varepsilon_0}\right) +\frac{\varepsilon_0}{r} e^{-\frac{a^2 + r^2}{2\varepsilon_0^2}} \sinh\left( \frac{r a}{\varepsilon_0^2}\right)     -\frac{\sqrt{2\pi}}{4} \left[ \erf\left( \frac{a-r}{\sqrt{2}\varepsilon_0}\right)+ \erf\left( \frac{a+r}{\sqrt{2}\varepsilon_0}\right) \right] \right\}  
   & r < a
\end{cases} \label{eq:3dpressure}
\end{align}
}
  The pressure profile from eq. \ref{eq:3dpressure} and the derivative of the pressure from eq. \ref{eq:dpdrfinal} are plotted in fig. \ref{fig:3dplot} for two values of $a/\varepsilon$. As discussed before, the  radius of support for the gradient of the pressure is $3.5 \varepsilon$, and for $a/\varepsilon =e$, the pressure difference $P_{\varepsilon, in}-P_{\varepsilon, out}$ varies from the Young-Laplace law.

We can now consider the value of the pressure at $r= 0$. For a sphere, the pressure jump should be $P(0) - P(\infty) = \frac{2\sigma}{a}$ in the limit $a/\varepsilon \rightarrow \infty$.
\begin{align}
&\lim_{r\rightarrow 0} -(s_{\alpha \alpha}-s_{\alpha \beta}) 4 \pi \varepsilon^5  \left\{ - \frac{2\varepsilon}{a} e^{-\frac{a^2}{\varepsilon^2}} \sinh\left( \frac{a^2}{\varepsilon^2}\right) +\frac{\sqrt{2\pi}}{2} \erf\left( \frac{2a}{\sqrt{2}\varepsilon}\right) +\frac{\varepsilon}{r} e^{-\frac{a^2 + r^2}{2\varepsilon^2}} \sinh\left( \frac{r a}{\varepsilon^2}\right)   \right. \nonumber \\
&  \quad\quad\quad\quad  \left.  -\frac{\sqrt{2\pi}}{4} \left[ \erf\left( \frac{a-r}{\sqrt{2}\varepsilon}\right)+ \erf\left( \frac{a+r}{\sqrt{2}\varepsilon}\right) \right] \right\} \nonumber \\
 &= -(s_{\alpha \alpha}-s_{\alpha \beta}) 4 \pi \varepsilon^5  \left\{ - \frac{\varepsilon}{a} \left(1-e^{-\frac{2a^2}{\varepsilon^2}} \right) +\frac{\sqrt{2\pi}}{2} \erf\left( \frac{2a}{\sqrt{2}\varepsilon}\right) + \frac{ a}{\varepsilon} e^{-\frac{a^2}{2\varepsilon^2}}-\frac{\sqrt{2\pi}}{2} \erf\left( \frac{a}{\sqrt{2}\varepsilon}\right)\right\} \label{eq:thirteen}
\end{align}
 Now, let $\eta = \frac{a}{\varepsilon}$ and expand the terms in eq. \ref{eq:thirteen} containing an error function at $\eta = \infty$. 
  \begin{align}
&-(s_{\alpha \alpha}-s_{\alpha \beta}) 4 \pi \varepsilon^5  \left\{ - \frac{1}{\eta} +\frac{1}{\eta} e^{-2\eta^2}+\frac{\sqrt{2\pi}}{2} \left( 1 + e^{-2\eta^2} \left[ -\frac{1}{\sqrt{2\pi}\eta} +\frac{1}{4\sqrt{2\pi}\eta^3}\right ]
\right)\right.\nonumber \\
&\quad \left. + \eta e^{-\frac{\eta^2}{2}}-\frac{\sqrt{2\pi}}{2} \left( 
1 + e^{-\eta^2/2} \left[ -\frac{\sqrt{2}}{\sqrt{\pi}\eta} +\frac{\sqrt{2}}{\sqrt{\pi}\eta^3}\right ] \right) + \mathcal{O} \left(e^{-\eta^2/2}\frac{1}{\eta^5} \right) \right\} \nonumber \\
&=(s_{\alpha \alpha}-s_{\alpha \beta}) 4 \pi \varepsilon^5  \left( \frac{1}{\eta}\right)+ \mathcal{O} \left(e^{-\eta^2/2}\eta \right)  \nonumber \\
&=(s_{\alpha \alpha}-s_{\alpha \beta}) \frac{4 \pi \varepsilon^6}{a}+ \mathcal{O} \left(e^{-\eta^2/2}\eta \right) 
\end{align}
 Therefore up to order $\mathcal{O} \left(e^{-\eta^2/2}\eta \right) $, 
 \begin{align}
 P(0) - P(\infty)\approx (s_{\alpha \alpha}-s_{\alpha \beta}) \frac{4 \pi \varepsilon^6}{a}-A(s_{\alpha \alpha}-s_{\alpha \beta}) \frac{4 \pi \varepsilon_0^6}{a}  &=  \frac{\sigma}{2\lambda}\frac{4 \pi}{a}\left(\varepsilon^6-A\varepsilon_0^6\right) \nonumber \\
  &=  \frac{\sigma}{2\pi\left(\varepsilon^6-A\varepsilon_0^6\right)}\frac{4 \pi}{a}\left(\varepsilon^6-A\varepsilon_0^6\right)  \nonumber\\
    &=   \frac{2\sigma}{a}
 \end{align}
 
We now consider the general case, where $\theta_i$ can vary. 
At equilibrium for a spherical symmetrical system we get
\begin{equation}
\frac{dp}{dr} = \int_{0}^{2 \pi} \int_{0}^{\pi} \int_{0}^{\8} f(r,s) s^2 \cos \phi \sin \phi ds d \phi d \theta,
\end{equation} 
where we choose the positive $z$-axis as the radial direction without loss of generality. The domain $\Omega_\alpha$ is taken as the ball of radius $a$ 
centered at the origin, and the local spherical coordinate system for the integration is centered at $r$, and thus
the factor $-\cos \phi$ arises from the quotient $(\vecx - \vecy)/|\vecx - \vecy|$ in the local coordinate system. 
We will compute the pressure profile in the radial direction of the ball. The domain outside the ball will be denoted as $\Omega_\beta$. The open domain inside the ball is denoted as $\Omega_\alpha$. 
Notice that if the function $f(r,s)$ is angular symmetric and compactly supported, which is the case for pairwise nonlocal
interactions considered in this study, then 
$$ \int_{0}^{2 \pi} \int_{0}^{\pi} \int_{0}^{\8} f(r,s) s^2 \cos \phi \sin \phi ds d \phi d \theta =0.$$
Below we will denote $f(r, s)$ with coefficient $s_{\alpha\alpha},s_{\beta\beta}$, or $s_{\alpha\beta}$ as $f_{11},f_{22}$ and $f_{12}$, respectively. 

For $r \ge a$ we have
\begin{align}
\int_{0}^{2 \pi} \int_{0}^{\pi} \int_{0}^{\8} f(r,s) s^2 \cos \phi \sin \phi ds d \phi d \theta & =  
\int_{0}^{2 \pi} \int_{\pi-\arcsin \frac{a}{r}}^{\pi} \cos \phi \sin \phi \int_{s_l}^{s_u} (f_{12} - f_{11}) s^2 ds d \phi d \theta, 
\label{eqn:integ_r>a}
\end{align}
where the lower and upper bounds of integral are
$ s_{l} =-r \cos \phi - \sqrt{a^2 - r^2 \sin^2 \phi}$ and $s_u =-r \cos \phi + \sqrt{a^2 - r^2 \sin^2 \phi}.$ We consider a generic term in the integral of eq. \ref{eqn:integ_r>a}:
\begin{align}
\int_{s_l}^{s_u} k e^{-\frac{s^2}{2 \varepsilon^2}} s^3 ds & = \frac{k}{2} \int_{s_l}^{s_u} e^{-\frac{s^2}{2 \varepsilon^2}} s^2 ds^2 \nonumber \\
& = \frac{k}{2} \int_{s_l^2}^{s_u^2} e^{-\frac{x}{2 \varepsilon^2}} x dx \nonumber \\ 
& =\frac{k}{2} \cdot -2\varepsilon^2 e^{-\frac{x}{2 \varepsilon^2}} (2 \varepsilon^2 + x) \left. \right |_{s_l^2}^{s_u^2} \nonumber \\
& =-k \varepsilon^2 e^{-\frac{x}{2 \varepsilon^2}} (2 \varepsilon^2 + x) \left. \right |_{s_l^2}^{s_u^2},  \label{eqn:integ_r>a_inner}
\end{align}
which gives rise to two terms of the same form, one for $s_u^2$ and the other for $s_l^2$. We consider the integration of term for
$s_u^2$ with respect to $\phi$:
\begin{align}
I_o(k, \varepsilon, s_u) = & -\int_{\pi-\arcsin \frac{a}{r}}^{\pi} \cos \phi \sin \phi k \varepsilon^2 e^{-\frac{s_u^2}{2 \varepsilon^2}} (2 \varepsilon^2 + s_u^2) \;  d \phi  
\nonumber \\
= &-k \varepsilon^2 \int_{\pi - \arcsin \frac{a}{r}}^{\pi} \cos \phi \sin \phi 
  \displaystyle{e^{-\frac{r^2 \cos^2 \phi + a^2 - r^2 \sin^2 \phi - 2 r \cos \phi \sqrt{a^2 - r^2 \sin^2 \phi}}{2 \varepsilon^2}}} 
 \nonumber  \\
  &   \qquad  \left ( 2 \varepsilon^2 + r^2 \cos^2 \phi + a^2 - r^2 \sin^2 \phi - 2 r \cos \phi \sqrt{a^2 - r^2 \sin^2 \phi} \right )  \; 
d \phi \nonumber \\
= & k \varepsilon^2 \int_{-\sqrt{1-a^2/r^2}}^{-1} x
\displaystyle{e^{-\frac{r^2 x^2 + a^2 - r^2 (1-x^2) - 2 r x \sqrt{a^2 - r^2 (1-x^2)}}{2 \varepsilon^2}}}  \nonumber  \\
  &   \qquad \left ( 2 \varepsilon^2 + r^2 x^2 + a^2 - r^2 (1-x^2) - 2 r x \sqrt{a^2 - r^2 (1-x^2)} \right ) 
d x \nonumber \\
= & k \varepsilon^2 \int_{\sqrt{1-a^2/r^2}}^{1} x 
\displaystyle{e^{-\frac{r^2 x^2 + a^2 - r^2 (1-x^2) + 2 r x \sqrt{a^2 - r^2 (1-x^2)}}{2 \varepsilon^2}}} \nonumber  \\
  &   \qquad \left ( 2 \varepsilon^2 + r^2 x^2 + a^2 - r^2 (1-x^2) + 2 r x \sqrt{a^2 - r^2 (1-x^2)} \right )  \; 
d x \nonumber \\
= & \frac{k \varepsilon^2}{r^2} \int_{\sqrt{D}}^r x 
\displaystyle{e^{-\frac{2x^2 -D + 2 x \sqrt{x^2-D}}{2 \varepsilon^2}}}
  \left ( 2 \varepsilon^2 + 2x^2 - D + 2 x \sqrt{x^2-D} \right ) \;  d x,
\end{align}
where $D = r^2-a^2$. Similarly, we have 
\begin{align}
I_o(k,\varepsilon, s_l) = & -\int_{\pi-\arcsin \frac{a}{r}}^{\pi} \cos \phi \sin \phi k \varepsilon^2 e^{-\frac{s_l^2}{2 \varepsilon^2}} (2 \varepsilon^2 + s_l^2)  \; d \phi  
\nonumber \\
= & \frac{k \varepsilon^2}{r^2} \int_{\sqrt{D}}^r x 
\displaystyle{e^{-\frac{2x^2 - D - 2 x \sqrt{x^2-D}}{2 \varepsilon^2}}}
  \left ( 2 \varepsilon^2 + 2x^2 - D - 2 x \sqrt{x^2-D} \right )  \; d x.
\end{align}
The full integral in eq. \ref{eqn:integ_r>a} is now equal to 
\begin{align}
\int_{0}^{2 \pi} \int_{\pi - \arcsin \frac{a}{r}}^{\pi} \cos \phi \sin \phi& \int_{s_l}^{s_u} (f_{12} - f_{11}) s^2 ds d \phi d \theta = 2 \pi \left[  I_o(-A s_{\alpha\beta}, \varepsilon_0, s_u)  +
I_o(A s_{\alpha\beta}, \varepsilon_0, s_l) + I_o(- s_{\alpha\beta},\varepsilon,s_l) \right. \nonumber \\
&\left. + I_o(s_{\alpha\beta},\varepsilon,s_u) +I_o(A  s_{\alpha\alpha}, \varepsilon_0, s_u) + I_o(- s_{\alpha\alpha},\varepsilon,s_u) +  I_o(-A s_{\alpha\alpha}, \varepsilon_0, s_l) + I_o(s_{\alpha\alpha},\varepsilon,s_l) \right].  \label{eqn:integ_r>a_2}
\end{align}
For $r < a$ we have
\begin{align}
\int_{0}^{2 \pi} \int_{0}^{\pi} \int_{0}^{\8} f(r,s) s^2 \cos \phi \sin \phi ds d \phi d \theta & =  
\int_{0}^{2 \pi} \int_{0}^{\pi} \cos \phi \sin \phi \int_{0}^{s_u} (f_{22} - f_{12}) s^2 ds d \phi d \theta, 
\label{eqn:integ_r<a}
\end{align}
where the upper bound of the integral is
$ s_{u} = \sqrt{a^2 - r^2 \sin \phi^2} - r \cos \phi.$
We consider the following generic term in the integral of eq. \ref{eqn:integ_r<a}:
\begin{align}
\int_{0}^{s_u} k e^{-\frac{s^2}{2 \varepsilon^2}} s^3 ds  = 
\frac{k}{2} \int_{s_l}^{s_u} e^{-\frac{s^2}{2 \varepsilon^2}} s^2 ds^2& = \frac{k}{2} \int_{0}^{s_u^2} e^{-\frac{x}{2 \varepsilon^2}} x dx =\left[ -k \varepsilon^2 e^{-\frac{x}{2 \varepsilon^2}} (2 \varepsilon^2 + x)  \right]_{0}^{s_u^2}.  
\end{align}
This gives rise to two terms. For the upper bound $s_u$ we compute
\begin{align}
I_i(k, \varepsilon, s_u) = 
& -k \varepsilon^2 \int_{0}^{\pi} \cos \phi \sin \phi 
e^{-\frac{ a^2 - r^2 \sin^2 \phi + r^2 \cos^2 \phi - 2 r \cos \phi \sqrt{a^2 - r^2 \sin^2 \phi}}{2 \varepsilon^2}}    \nonumber \\
& \qquad \left ( 2 \varepsilon^2 + a^2 - r^2 \sin^2 \phi + r^2 \cos^2 \phi - 2 r \cos \phi \sqrt{a^2 - r^2 \sin^2 \phi} \right ) \; d \phi  
\nonumber \\
= & k \varepsilon^2 \int_{0}^{\pi} \cos \phi  
e^{-\frac{ a^2 - r^2 \sin^2 \phi + r^2 \cos^2 \phi - 2 r \cos \phi \sqrt{a^2 - r^2 \sin^2 \phi}}{2 \varepsilon^2}}    \nonumber \\
& \qquad \left ( 2 \varepsilon^2 + a^2 - r^2 \sin^2 \phi + r^2 \cos^2 \phi - 2 r \cos \phi \sqrt{a^2 - r^2 \sin^2 \phi} \right )\; d (\cos \phi)  
\nonumber \\
= &-k \varepsilon^2 \int_{-1}^{1} x 
e^{-\frac{ a^2 - r^2 + 2r^2 x^2 - 2 r x \sqrt{a^2 - r^2 + r^2 x^2}}{2 \varepsilon^2}}  \left ( 2 \varepsilon^2 + a^2 - r^2 + 2r^2 x^2 - 2 r x \sqrt{a^2 - r^2 + r^2 x^2} \right ) \;d x. 
\end{align}
If $r \ne 0$, then 
\begin{align}
I_i(k, \varepsilon, s_u)  
= &-\frac{k \varepsilon^2}{r^2} \int_{-1}^{1} r x 
e^{-\frac{ a^2 - r^2 + 2r^2 x^2 - 2 r x \sqrt{a^2 - r^2 + r^2 x^2}}{2 \varepsilon^2}} \left ( 2 \varepsilon^2 + a^2 - r^2 + 2r^2 x^2 - 2 r x \sqrt{a^2 - r^2 + r^2 x^2} \right ) d (rx)  
\nonumber \\
= &-\frac{k \varepsilon^2}{r^2} \int_{-r}^{r} x 
e^{-\frac{ a^2 - r^2 + 2 x^2 - 2 x \sqrt{a^2 - r^2 + x^2}}{2 \varepsilon^2}}   \left ( 2 \varepsilon^2 + a^2 - r^2 + 2 x^2 - 2 x \sqrt{a^2 - r^2 + x^2} \right ) d x, 
\end{align}
otherwise,
\begin{align}
I_i(k, \varepsilon, s_u)  = 
-k \varepsilon^2 \int_{-1}^{1} x e^{-\frac{a^2}{2 \varepsilon^2}} \cdot ( 2 \varepsilon^2 + a^2) d x = 0. 
\end{align}
For the lower bound $0$ we have 
\begin{align}
I_i(k, \varepsilon, 0) = & -k \varepsilon^2 \int_{0}^{\pi} \cos \phi \sin \phi  (2 \varepsilon^2) d \phi = 0.
\end{align}
The full integration in eq. \ref{eqn:integ_r<a} gives:
\begin{align}
\int_{0}^{2 \pi} \int_{0}^{\pi} \cos \phi \sin \phi \int_{0}^{s_u} (f_{22} - f_{12}) s^2 ds d \phi d \theta   
 = & 2 \pi \big (  I_i(-A s_{\beta \beta}, \varepsilon_0, s_u) + I_i(s_{\beta \beta},\varepsilon,s_u) + I_i(A s_{\alpha \beta}, \varepsilon_0, s_u) + I_i(-s_{\alpha \beta},\varepsilon,s_u) \big ). \label{eqn:integ_r<a_2}
\end{align}

\section{Two-dimensional flat interface}
\label{sec:2dflat}

We consider a flat interface located at $x = 0$. Note that in this case, the coefficient $s(\bx, \bx_i) = s(x, x_i)$ depends only on the $x-$coordinates. We start with:
\begin{equation}
\nabla P(\bx) = -\int s(\bx, \by) f_\varepsilon(\bx - \by)\frac{\bx-\by}{|\bx-\by|} \; d\by = 
- \int s(\bx, \by) \left[e^{\frac{-|\bx-\by|^2}{2\varepsilon^2}}-Ae^{\frac{-|\bx-\by|^2}{2\varepsilon_0^2}}\right](\bx-\by) \; d\by 
\end{equation}

First, consider
\begin{align}
\frac{\partial P}{\partial y_i} &=
- \int_{-\infty}^\infty \int_{-\infty}^\infty s(x, x_i)  \left[e^{\frac{-(x-x_i)^2+(y-y_i)^2}{2\varepsilon^2}}-Ae^{\frac{-(x-x_i)^2+(y-y_i)^2}{2\varepsilon_0^2}}\right](y-y_i) \; dy dx\nonumber \\
&= 0
\end{align}
using the fact that 
$ \int_{-\infty}^\infty e^{\frac{-(y-y_i)^2}{2\varepsilon^2}}(y-y_i) \; dy dx $.
 Now, 
 \begin{align}
\frac{\partial P}{\partial x_i} &=
- \int_{-\infty}^\infty \int_{-\infty}^\infty s(x, x_i)  \left[-e^{\frac{(x-x_i)^2+(y-y_i)^2}{2\varepsilon^2}}-Ae^{-\frac{(x-x_i)^2+(y-y_i)^2}{2\varepsilon_0^2}}\right](x-x_i) \; dy dx \label{eq:twenty}
\end{align}
Consider one integral from eq. \ref{eq:twenty}: 
  \begin{align}
\int_{-\infty}^\infty \int_{-\infty}^\infty s(x, x_i)  e^{-\frac{(x-x_i)^2+(y-y_i)^2}{2\varepsilon^2}}(x-x_i) \; dy dx &=\int_{-\infty}^\infty s(x, x_i)  e^{\frac{-(x-x_i)^2}{2\varepsilon^2}}(x-x_i)  \int_{-\infty}^\infty  e^{\frac{-(y-y_i)^2}{2\varepsilon^2}} \; dy dx  \nonumber \\
&=\sqrt{2 \pi} \varepsilon \int_{-\infty}^\infty s(x, x_i)  e^{\frac{-(x-x_i)^2}{2\varepsilon^2}}(x-x_i)  \; dx
\end{align}
Then, for $x_i < 0$, 
\begin{align}
\sqrt{2 \pi} \varepsilon \int_{-\infty}^\infty s(x, x_i)  e^{\frac{-(x-x_i)^2}{2\varepsilon^2}}(x-x_i)  \; dx&=\sqrt{2 \pi} \varepsilon \int_{-\infty}^0 s_{\alpha \alpha} e^{\frac{-(x-x_i)^2}{2\varepsilon^2}}(x-x_i)  \; dx  + \sqrt{2 \pi} \varepsilon \int_{0}^\infty s_{\alpha \beta}  e^{\frac{-(x-x_i)^2}{2\varepsilon^2}}(x-x_i)  \; dx  \nonumber \\
&=- \sqrt{2 \pi} \varepsilon^3 s_{\alpha \alpha} e^{\frac{-x_i^2}{2\varepsilon^2}} + \sqrt{2 \pi} \varepsilon^3 s_{\alpha \beta} e^{\frac{-x_i^2}{2\varepsilon^2}} \nonumber \\
&=- \sqrt{2 \pi} \varepsilon^3 e^{\frac{-x_i^2}{2\varepsilon^2}} \left[ s_{\alpha \alpha} -  s_{\alpha \beta} \right]
\end{align}
 Therefore, for $x_i < 0$, eq. \ref{eq:twenty} gives:
  \begin{align}
\frac{\partial P}{\partial x_i} &=
\sqrt{2 \pi} \varepsilon^3 e^{\frac{-x_i^2}{2\varepsilon^2}} \left[ s_{\alpha \alpha} -  s_{\alpha \beta} \right]-A \sqrt{2 \pi} \varepsilon_0^3 e^{\frac{-x_i^2}{2\varepsilon_0^2}} \left[ s_{\alpha \alpha} -  s_{\alpha \beta} \right] \nonumber \\
&=
\sqrt{2 \pi}\left[ s_{\alpha \alpha} -  s_{\alpha \beta} \right] \left( \varepsilon^3 e^{\frac{-x_i^2}{2\varepsilon^2}} -A  \varepsilon_0^3 e^{\frac{-x_i^2}{2\varepsilon_0^2}} \right) \label{eq:twentytwo}
\end{align}
For $x_i \geq 0$, the solution will be the same as eq. \ref{eq:twentytwo} with $s_{\alpha \beta} $ and $s_{\alpha \alpha}$ switched. 

Therefore, if $x < 0$, 
\begin{align}
P(x) - P(-\infty) &= \int_{-\infty}^x \sqrt{2 \pi}\left[ s_{\alpha \alpha} -  s_{\alpha \beta} \right] \left( \varepsilon^3 e^{\frac{-x_i^2}{2\varepsilon^2}} -A  \varepsilon_0^3 e^{\frac{-x_i^2}{2\varepsilon_0^2}} \right) \; dx_i\nonumber \\
&= \sqrt{2 \pi}\left[ s_{\alpha \alpha} -  s_{\alpha \beta} \right] \left[ 
\frac{\sqrt{2\pi}\varepsilon^4}{2}\erf\left(\frac{x_i}{\sqrt{2}\varepsilon} \right)-A\frac{\sqrt{2\pi}\varepsilon_0^4}{2}\erf\left(\frac{x_i}{\sqrt{2}\varepsilon_0} \right)
\right]_{-\infty}^x \nonumber \\
&= {\pi}\left[ s_{\alpha \alpha} -  s_{\alpha \beta} \right] \left[ 
\varepsilon^4 \left(\erf\left(\frac{x}{\sqrt{2}\varepsilon} \right)+1\right)-A\varepsilon_0^4\left(\erf\left(\frac{x}{\sqrt{2}\varepsilon_0} \right)+1\right)
\right]
\end{align}
 and for $x \geq 0$:
 \begin{align}
P(x) - P(-\infty) &= \int_{-\infty}^0 \sqrt{2 \pi}\left[ s_{\alpha \alpha} -  s_{\alpha \beta} \right] \left( \varepsilon^3 e^{\frac{-x_i^2}{2\varepsilon^2}} -A  \varepsilon_0^3 e^{\frac{-x_i^2}{2\varepsilon_0^2}} \right) \; dx_i -\int_{0}^x \sqrt{2 \pi}\left[ s_{\alpha \alpha} -  s_{\alpha \beta} \right] \left( \varepsilon^3 e^{\frac{-x_i^2}{2\varepsilon^2}} -A  \varepsilon_0^3 e^{\frac{-x_i^2}{2\varepsilon_0^2}} \right) \; dx_i \nonumber \\
&=  \pi\left[ s_{\alpha \alpha} -  s_{\alpha \beta} \right] \left[ 
\varepsilon^4-A\varepsilon_0^4
\right]-{2 \pi}\left[ s_{\alpha \alpha} -  s_{\alpha \beta}\right]  \left[ 
\frac{\varepsilon^4}{2}\erf\left(\frac{x_i}{\sqrt{2}\varepsilon} \right)-A\frac{\varepsilon_0^4}{2}\erf\left(\frac{x_i}{\sqrt{2}\varepsilon_0} \right)
\right]_{0}^x \nonumber \\
&=-{\pi}\left[ s_{\alpha \alpha} -  s_{\alpha \beta}\right]  \left[ 
\varepsilon^4\left(\erf\left(\frac{x}{\sqrt{2}\varepsilon} \right)-1\right)-A\varepsilon_0^4\left(\erf\left(\frac{x}{\sqrt{2}\varepsilon_0} \right)-1\right)
\right]
\end{align}
   
  \begin{figure}[ht]
\centering
    \includegraphics[width=0.45\linewidth]{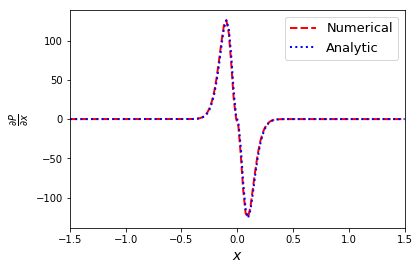}
        \includegraphics[width=0.45\linewidth]{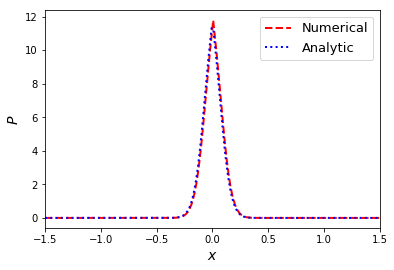}
    \caption{(Left) Comparison of numerical integration of eq. \ref{eq:twenty} and the analytic solution in eq. \ref{eq:dpdx_2d_plane} with $\varepsilon = 1/10$. (Right) Comparison of the analytic solution in eq. \ref{eq:P_2d_flat} and numerically integrating eq.  \ref{eq:dpdx_2d_plane}.}
    \label{fig:2d_plane}
\end{figure}

To summarize the results for a flat interface in two dimensions, 
\boxalign{
\begin{align}
\frac{\partial P}{\partial x_i}&=
\begin{cases} \sqrt{2 \pi}\left[ s_{\alpha \alpha} -  s_{\alpha \beta} \right] \left( \varepsilon^3 e^{-\frac{x_i^2}{2\varepsilon^2}} -A  \varepsilon_0^3 e^{-\frac{x_i^2}{2\varepsilon_0^2}} \right), & x_i < 0 \\
 -\sqrt{2 \pi}\left[ s_{\alpha \alpha} -  s_{\alpha \beta} \right] \left( \varepsilon^3 e^{-\frac{x_i^2}{2\varepsilon^2}} -A  \varepsilon_0^3 e^{-\frac{x_i^2}{2\varepsilon_0^2}} \right),  & x_i \geq 0 \\
\end{cases}   \label{eq:dpdx_2d_plane} \\
\frac{\partial P}{\partial y_i} &= 0 \\
 P(x) - P(-\infty) &= \begin{cases}
 {\pi}\left[ s_{\alpha \alpha} -  s_{\alpha \beta} \right] \left[ 
\varepsilon^4 \left(\erf\left(\frac{x}{\sqrt{2}\varepsilon} \right)+1\right)-A\varepsilon_0^4\left(\erf\left(\frac{x}{\sqrt{2}\varepsilon_0} \right)+1\right)
\right], & x < 0 \label{eq:P_2d_flat} \\  
-{\pi}\left[ s_{\alpha \alpha} -  s_{\alpha \beta}\right]  \left[ 
\varepsilon^4\left(\erf\left(\frac{x}{\sqrt{2}\varepsilon} \right)-1\right)-A\varepsilon_0^4\left(\erf\left(\frac{x}{\sqrt{2}\varepsilon_0} \right)-1\right)
\right], & x \geq 0
\end{cases} 
\end{align}
}The numerical and analytic solutions are compared in fig. \ref{fig:2d_plane}. The support for both the gradient of the pressure and the pressure profiles is $3.5 \varepsilon$, as for the case of a circle or sphere. The pressure profile is similar to that seen in MD simulations \citep{Masuda2011, Marchand2011} and Smoothed Particle Hydrodynamics \citep{Tartakovsky2016}.

\section{Three-dimensional flat interface}

We again consider a flat interface located at $x = 0$.

First, consider
\begin{align}
\frac{\partial P}{\partial y_i} &=
- \int_{-\infty}^\infty \int_{-\infty}^\infty  \int_{-\infty}^\infty s(x, x_i)  \left[e^{\frac{-(x-x_i)^2+(y-y_i)^2+(z-z_i)^2}{2\varepsilon^2}}-Ae^{\frac{-(x-x_i)^2+(y-y_i)^2+(z-z_i)^2}{2\varepsilon_0^2}}\right](y-y_i) \; dy dz dx\nonumber \\
&= 0
\end{align}
using the fact that 
$ \int_{-\infty}^\infty e^{\frac{-(y-y_i)^2}{2\varepsilon^2}}(y-y_i) \; dy dx $.
Similarly, 
\begin{align}
\frac{\partial P}{\partial z_i} &=
- \int_{-\infty}^\infty \int_{-\infty}^\infty  \int_{-\infty}^\infty s(x, x_i)  \left[e^{\frac{-(x-x_i)^2+(y-y_i)^2+(z-z_i)^2}{2\varepsilon^2}}-Ae^{\frac{-(x-x_i)^2+(y-y_i)^2+(z-z_i)^2}{2\varepsilon_0^2}}\right](z-z_i) \; dz dy dx\nonumber \\
&= 0.
\end{align}

 Now, 
 \begin{align}
\frac{\partial P}{\partial x_i} &=
- \int_{-\infty}^\infty \int_{-\infty}^\infty  \int_{-\infty}^\infty s(x, x_i)  \left[e^{\frac{-(x-x_i)^2+(y-y_i)^2+(z-z_i)^2}{2\varepsilon^2}}-Ae^{\frac{-(x-x_i)^2+(y-y_i)^2+(z-z_i)^2}{2\varepsilon_0^2}}\right](x-x_i) \; dz dy dx\label{eq:twentyone}
\end{align}
Using the same argument as in sec. \ref{sec:2dflat}, for $x_i < 0$, we get:
  \begin{align}
\frac{\partial P}{\partial x_i} 
&=
{2 \pi}\left[ s_{\alpha \alpha} -  s_{\alpha \beta} \right] \left( \varepsilon^4 e^{\frac{-x_i^2}{2\varepsilon^2}} -A  \varepsilon_0^4 e^{\frac{-x_i^2}{2\varepsilon_0^2}} \right) 
\end{align}
For $x_i \geq 0$, the solution will be the same as eq. \ref{eq:dpdx_3d_plane} with $s_{\alpha \beta} $ and $s_{\alpha \alpha}$ switched. 

Therefore, if $x < 0$, 
\begin{align}
P(x) - P(-\infty) 
&= \sqrt{2\pi^3}\left[ s_{\alpha \alpha} -  s_{\alpha \beta} \right] \left[ 
\varepsilon^5 \left(\erf\left(\frac{x}{\sqrt{2}\varepsilon} \right)+1\right)-A\varepsilon_0^5\left(\erf\left(\frac{x}{\sqrt{2}\varepsilon_0} \right)+1\right)
\right]\label{eq:P_3d_flata}
\end{align}
 and for $x \geq 0$:
 \begin{align}
P(x) - P(-\infty) &=- \sqrt{2\pi^3}\left[ s_{\alpha \alpha} -  s_{\alpha \beta}\right]  \left[ 
\varepsilon^5\left(\erf\left(\frac{x}{\sqrt{2}\varepsilon} \right)-1\right)-A\varepsilon_0^5\left(\erf\left(\frac{x}{\sqrt{2}\varepsilon_0} \right)-1\right)
\right]\label{eq:P_3d_flatb}
\end{align}

 \begin{figure}[ht]
\centering
    \includegraphics[width=0.45\linewidth]{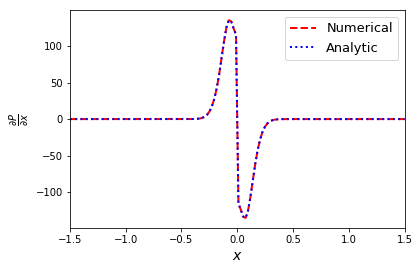}
        \includegraphics[width=0.45\linewidth]{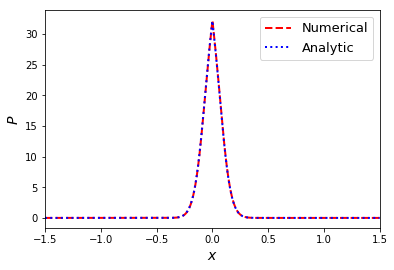}
    \caption{(Left) Comparison of numerical integration of eq. \ref{eq:twentyone} and the analytic solution in eq. \ref{eq:dpdx_3d_plane} with $\varepsilon = 1/10$. (Right) Comparison of the analytic solution in eq. \ref{eq:P_3d_flat} and numerically integrating eq.  \ref{eq:dpdx_3d_plane}.}
    \label{fig:3d_plane}
\end{figure}

 Summarizing these results,
\boxalign{
\begin{align}
\frac{\partial P}{\partial x_i}&=
\begin{cases} {2 \pi}\left[ s_{\alpha \alpha} -  s_{\alpha \beta} \right] \left( \varepsilon^4 e^{-\frac{x_i^2}{2\varepsilon^2}} -A  \varepsilon_0^4 e^{-\frac{x_i^2}{2\varepsilon_0^2}} \right) , & x_i < 0 \\
 -{2 \pi}\left[ s_{\alpha \alpha} -  s_{\alpha \beta} \right] \left( \varepsilon^4 e^{-\frac{x_i^2}{2\varepsilon^2}} -A  \varepsilon_0^4 e^{-\frac{x_i^2}{2\varepsilon_0^2}} \right) ,  & x_i \geq 0 \\
\end{cases}   \label{eq:dpdx_3d_plane} \\
\frac{\partial P}{\partial y_i} &= 0 \\
\frac{\partial P}{\partial z_i} &= 0 \\
 P(x) - P(-\infty) &= \begin{cases}
 \sqrt{2\pi^3}\left[ s_{\alpha \alpha} -  s_{\alpha \beta} \right] \left[ 
\varepsilon^5 \left(\erf\left(\frac{x}{\sqrt{2}\varepsilon} \right)+1\right)-A\varepsilon_0^5\left(\erf\left(\frac{x}{\sqrt{2}\varepsilon_0} \right)+1\right)
\right], & x < 0 \label{eq:P_3d_flat} \\  
-\sqrt{2\pi^3}\left[ s_{\alpha \alpha} -  s_{\alpha \beta}\right]  \left[ 
\varepsilon^5\left(\erf\left(\frac{x}{\sqrt{2}\varepsilon} \right)-1\right)-A\varepsilon_0^5\left(\erf\left(\frac{x}{\sqrt{2}\varepsilon_0} \right)-1\right)
\right], & x \geq 0
\end{cases}
\end{align}
} The numerical and analytic solutions are compared in fig. \ref{fig:3d_plane}.


\section{Pressure based on the Lennard-Jones potentials}

In this section we consider an analytical solution with $f_\varepsilon$ in eqs. \ref{eq:Nonlocal_force}, \ref{eq:forceshape}, with the Lennard-Jones (LJ) forces. We include a radial distribution function in the integral to give: 
\begin{equation}
\nabla P(\bx)= - \int_\Omega g(\bx, \by) F_{\alpha \beta}^{int, LJ}(\bx, \by) \; d\by \label{eq:LJforce}.
\end{equation}
The 9-6 LJ potential is given by 
\begin{equation}
U_{jk} = \frac{27\varepsilon_{jk}}{4}\left[ \left(\frac{\sigma_{jk}}{|\bx|}\right)^9-\left(\frac{\sigma_{jk}}{|\bx|}\right)^6\right]
\end{equation} where $j$ and $k$ each denote a phase $\alpha$ or $\beta$. 
Note that 
\begin{equation}
\frac{\partial U_{jk}}{\partial x} = \frac{27\varepsilon_{jk}}{2\sigma}x\left[-9 \left(\frac{\sigma_{jk}}{|\bx|}\right)^{10}+6\left(\frac{\sigma_{jk}}{|\bx|}\right)^7\right]
\end{equation}
Then, 
\begin{equation}
\frac{\partial P}{\partial x_i} = \int_\Omega g(\bx_i, \bx_n) \frac{27\varepsilon_{jk}}{2\sigma_{jk}}(x_i-x_n)\left[-9 \left(\frac{\sigma_{jk}}{|\bx_i-\bx_n|}\right)^{10}+6\left(\frac{\sigma_{jk}}{|\bx_i-\bx_n|}\right)^7\right] \; d\bx_n 
\end{equation}
We will consider the case where the interface is an infinite plane located at $x = 0$, with domain $\Omega_{\alpha} = (-\infty, 0]\times (-\infty, \infty)$ and $\Omega_{\beta} = (0, \infty)\times (-\infty, \infty). $

\subsection{$g(x, y) = 1$:}
\begin{align}
\frac{\partial P}{\partial x_i} &= \int_\Omega \frac{27\varepsilon_{jk}}{2\sigma_{jk}}(x_i-x_n)\left[-9 \left(\frac{\sigma_{jk}}{|\bx_i-\bx_n|}\right)^{10}+6\left(\frac{\sigma_{jk}}{|\bx_i-\bx_n|}\right)^7\right] \; d\bx_n \nonumber \\
&= \int_{-\infty}^0\int_{-\infty}^\infty  \frac{27\varepsilon_{\alpha\alpha}}{2\sigma_{\alpha\alpha}}(x_i-x_n)\left[-9 \left(\frac{\sigma_{\alpha\alpha}}{|\bx_i-\bx_n|}\right)^{10}+6\left(\frac{\sigma_{\alpha\alpha}}{|\bx_i-\bx_n|}\right)^7\right] \; dy_n dx_n  \nonumber\\
&\quad +  \int_0^{\infty}\int_{-\infty}^\infty  \frac{27\varepsilon_{\alpha\beta}}{2\sigma_{\alpha\beta}}(x_i-x_n)\left[-9 \left(\frac{\sigma_{\alpha\beta}}{|\bx_i-\bx_n|}\right)^{10}+6\left(\frac{\sigma_{\alpha\beta}}{|\bx_i-\bx_n|}\right)^7\right] \; dy_n dx_n
\end{align}
Assume $y_i = 0$. Consider a representative integral: 
\begin{align}
&\int_{-\infty}^\infty  \frac{27\varepsilon_{\alpha\alpha}}{2\sigma_{\alpha\alpha}}(x_i-x_n)\left[-9 \left(\frac{\sigma_{\alpha\alpha}}{((x_i-x_n)^2+y_n^2)^{0.5}}\right)^{10}+6\left(\frac{\sigma_{\alpha\alpha}}{((x_i-x_n)^2+y_n^2)^{0.5}}\right)^7\right] \; dy_n  \nonumber \\
&= \frac{27\varepsilon_{\alpha\alpha}}{2\sigma_{\alpha\alpha}}(x_i-x_n)\left[-9\sigma_{\alpha\alpha}^{10} \frac{35 \pi}{128(x_i-x_n)^9}
+6\sigma_{\alpha\alpha}^7 \frac{16}{15(x_i-x_n)^6}\right]  \nonumber\\
&= \frac{27\varepsilon_{\alpha\alpha}}{2\sigma_{\alpha\alpha}} \left[-\sigma_{\alpha\alpha}^{10} \frac{315 \pi}{128(x_i-x_n)^8}
+\sigma_{\alpha\alpha}^7 \frac{32}{5(x_i-x_n)^5}\right]  
\end{align}
Therefore, 
\begin{align}
\frac{\partial P}{\partial x_i} &= \int_{-\infty}^0 \frac{27\varepsilon_{\alpha\alpha}}{2\sigma_{\alpha\alpha}} \left[-\sigma_{\alpha\alpha}^{10} \frac{315 \pi}{128(x_i-x_n)^8}
+\sigma_{\alpha\alpha}^7 \frac{32}{5(x_i-x_n)^5}\right] \; dx_n \nonumber \\
&\quad +  \int_0^{\infty} \frac{27\varepsilon_{\alpha\beta}}{2\sigma_{\alpha\beta}} \left[-\sigma_{\alpha\beta}^{10} \frac{315 \pi}{128(x_i-x_n)^8}
+\sigma_{\alpha\beta}^7 \frac{32}{5(x_i-x_n)^5}\right] \;  dx_n  \nonumber\\
 &=  \frac{27\varepsilon_{\alpha\alpha}\sigma_{\alpha\alpha}^6}{2} \left[-\sigma_{\alpha\alpha}^{3} \frac{45 \pi}{128x_i^7}
+\frac{8}{5x_i^4}\right] +  \frac{27\varepsilon_{\alpha\beta}\sigma_{\alpha\beta}^6}{2} \left[\sigma_{\alpha\beta}^{3} \frac{45 \pi}{128x_i^7}
-\frac{8}{5x_i^4}\right] 
\end{align}
And thus, 
\begin{align}
P(x)-P(-\infty) &= \int_{-\infty}^x \frac{27\varepsilon_{\alpha\alpha}\sigma_{\alpha\alpha}^6}{2} \left[-\sigma_{\alpha\alpha}^{3} \frac{45 \pi}{128x_i^7}
+\frac{8}{5x_i^4}\right] +  \frac{27\varepsilon_{\alpha\beta}\sigma_{\alpha\beta}^6}{2} \left[\sigma_{\alpha\beta}^{3} \frac{45 \pi}{128x_i^7} 
-\frac{8}{5x_i^4}\right] \nonumber \\
&= \frac{27\varepsilon_{\alpha\alpha}\sigma_{\alpha\alpha}^6}{2} \left[\sigma_{\alpha\alpha}^{3} \frac{5 \pi}{92x^6}
-\frac{8}{15x^3}\right] -  \frac{27\varepsilon_{\alpha\beta}\sigma_{\alpha\beta}^6}{2} \left[\sigma_{\alpha\beta}^{3} \frac{5 \pi}{92x^6}-
\frac{8}{15x^3}\right] 
\end{align}
This integral diverges as $x \rightarrow 0$. 

\subsection{$g(x, y) = \exp(-U(r)/kT)$:}

\begin{equation}
\frac{\partial P}{\partial x_i} = \int_\Omega e^{- \frac{27\varepsilon_{jk}}{4kT}\left[ \left(\frac{\sigma_{jk}}{|\bx_i-\bx_n|}\right)^9-\left(\frac{\sigma_{jk}}{|\bx_i-\bx_n|}\right)^6\right]}  \frac{27\varepsilon_{jk}}{2\sigma_{jk}}(x_i-x_n)\left[-9 \left(\frac{\sigma_{jk}}{|\bx_i-\bx_n|}\right)^{10}+6\left(\frac{\sigma_{jk}}{|\bx_i-\bx_n|}\right)^7\right] \; d\bx_n .
\end{equation}

Note that: 
\begin{align}
\frac{\partial P}{\partial x_i} &= \int_{-\infty}^\infty  \int_{-\infty}^0 e^{-\frac{U(|\bx_i-\bx_n|)}{kT}} \frac{\partial U(|\bx_i-\bx_n|)}{\partial x_n}  \; d x_n dy_n \nonumber  \\
&= \int_{-\infty}^\infty  \int_{V(-\infty)}^{V(0)} e^{-V} kT  \; d V dy_n \nonumber \\
&= \int_{-\infty}^\infty  \left[ -e^{-V} kT  \right]_{V(-\infty)}^{V(0)}\;dy_n \nonumber \\
&= \int_{-\infty}^\infty  \left[ -e^{-\frac{U(|\bx_i-\bx_n|)}{kT}} kT  \right]_{-\infty}^{0}\;dy_n\nonumber \\
&= -kT\int_{-\infty}^\infty  1- e^{-\frac{U\left(\sqrt{x_i^2 + (y_i-y_n)^2}\right)}{kT}}   \;dy_n \label{eq:eleven}
\end{align}
where $V = \frac{U}{kT}.$ The integral in eq. \ref{eq:eleven} will clearly diverge because of the $-kT\int_{-\infty}^\infty  1   \;dy_n$ term.

\section{Acknowledgements} 
This work was supported by the U.S. Department of Energy (DOE) Office of Science, Office of Advanced Scientific Computing Research as part of the New Dimension Reduction Methods and Scalable Algorithms for Nonlinear Phenomena project. Pacific Northwest National Laboratory is operated by Battelle for the DOE under Contract DE-AC05-76RL01830.

\bibliographystyle{abbrvnat}
\bibliography{biblio}

\end{document}